\documentclass[]{interact}

\usepackage{epstopdf}% To incorporate .eps illustrations using PDFLaTeX, etc.
\usepackage[caption=false]{subfig}% Support for small, `sub' figures and tables

\usepackage[numbers,sort&compress]{natbib}% Citation support using natbib.sty

\usepackage{multirow}
\bibpunct[, ]{[}{]}{,}{n}{,}{,}% Citation support using natbib.sty
\theoremstyle{plain}% Theorem-like structures provided by amsthm.sty

\theoremstyle{definition}

\theoremstyle{remark}

\begin{document}

\articletype{ARTICLE}% Specify the article type or omit as appropriate

\title{Monitoring Coefficient of Variation using One-Sided Run Rules control charts in the presence of Measurement Errors}

\author{
\name{Phuong Hanh Tran\textsuperscript{a,b}\thanks{Corresponding author. Phuong Hanh Tran. Email: tran@doct.uliege.be} and C\'edric Heuchenne\textsuperscript{b} and Huu Du Nguyen\textsuperscript{a}}
\affil{\textsuperscript{a}Division of Artificial Intelligence, Dong A University, Danang, Vietnam; \textsuperscript{b}HEC Li\`ege - Management School of the University of Li\`ege, Li\`ege, Belgium}
}

\maketitle

\begin{abstract}
We investigate in this paper the effect of the measurement error on the performance of Run Rules control charts monitoring the coefficient of variation (CV) squared. The previous Run Rules CV chart in the literature is improved slightly by monitoring the CV squared  using two one-sided Run Rules charts instead of monitoring the CV itself using a two-sided chart. The numerical results show that this improvement gives better performance in detecting process shifts. Moreover, we will show through simulation that the \textit{precision} and \textit{accuracy} errors do have negative effect on the performance of the proposed Run Rules charts. We also find out that taking multiple measurements per item is not an effective way to reduce these negative effects. 
\end{abstract}

\begin{keywords}
Run Rules chart, Markov chain, Coefficient of Variation, Measurement Errors.
\end{keywords}

\section{Introduction}

\label{sec:Intro}
Among important statistical characteristics of a variable, the coefficient of variation (CV) is widely used to evaluate the stability or concentration of the random variable around the mean. It is defined as the ratio between the standard deviation to the mean, $\gamma=\sigma/\mu$. In many industrial processes, keeping the value of this coefficient of a characteristic of interest within the permissible range means assuring the quality of products. A number of examples have been illustrated in the literature for the applications of the CV in industry. \cite{Castagliola_EWMA_CV_2011} presented an example where the quality of interest is the pressure test drop time from a sintering process manufacturing mechanical parts. In this example, the presence of a constant proportionality between the standard deviation of the pressure drop time and its mean was confirmed. The CV was then monitored to detect changes in the process variability. \cite{ye2018novel} showed that it is useful to monitor the CV in detecting the presence of chatter, a severe form of self-excited vibration in the machining process which leads to many machining problems. More examples about the need of using the CV as a measure of interest has been discussed in  \cite{Muhammada_2018_VSSEWMA_CV}. Because of its wide range of applications, monitoring the CV is a major objective in many studies in statistical process control, see, for example,  \cite{CastagliolaVSI_CV_2013},\cite{Castagliola2015_VSS_CV}, 
\cite{Yeong_VSI_CV_Direct_2017}, and \cite{Khaw_VSSI_CV_2017}.

Along with the development of the advanced control charts monitoring the CV with improved performance, recent researches are paying attention to the effect of the measurement error on the CV control chart. This makes these researches become more in touch reality since the measurement error is often present in practice. A Shewhart control chart monitoring the CV under the presence of measurement error (ME) was suggested by \cite{Yeong_CV_ME_2017}.  \cite{Tran_2018_CVME} improved the linear covariate error model for the CV in \cite{Yeong_CV_ME_2017} and then proposed the EWMA CV control chart with ME. Very recently, \cite{Nguyen_2019_VSICV_ME} and  \cite{tran2019performance} studied the effect of ME on the variable sampling interval control chart and the cumulative sum control chart monitoring the CV, respectively.

One of the reasons that leads to the introduction of many advanced control charts monitoring the CV is to overcome a drawback of the Shewhart CV chart that it is not sensitive to the large shifts. However, the Shewhart chart is still popularly used thanks to its simplicity in implementation. From this point of view, the Run Rules charts are advantageous: they are easy to implement (compared to, for example, the EWMA control chart or the CUSUM control chart, even these charts may bring better performance) and they can improve remarkably the performance of the Shewhart chart in detecting small or moderate process shifts. The aim of this paper is to investigate the performance of Run Rules CV control chart under the presence of ME. In fact, the Run Rules chart monitoring the CV has been studied in \cite{Castagliola_Run-Rules_CV_2013}. However, the ME has not been considered. Moreover, in this study the authors only focused on the two-sided charts (the one-sided chart has been mentioned, but quite sketchily without explanation for the design) with the CV monitored directly. We improve this design by monitoring the CV squared and presenting the design of the two one-sided Run Rules charts in detail.

The paper consists of eight sections and is organized as follows. Followed by the introduction in Section \ref{sec:Intro}, Section \ref{sec:CVdistribution} presents  a brief review of the distribution of the sample coefficient of variation. The design and the implementation of two one-sided Run Rules control charts monitoring the CV squared (denoted as RR$_{r,s}-{\gamma}^2$ charts) are presented in section \ref{sec: design no ME}. Section \ref{sec: perform no ME} is for the performance of these charts. A linear covariate error model for the CV is reintroduced in section \ref{sec: ME model}. The design of control charts in the presence of measurement errors and the effect of the measurement error on the RR$_{r,s}-{\gamma}^2$ charts are displayed in section \ref{sec:perform me}. Section \ref{sec: example} is devoted to an illustrative example. Some concluding remarks are given in section \ref{sec:conclu} to conclude.

%%%%%%%%%%%%%%%%%%%%%%%%%%%%%%%%%%%%%%%%%%%%%%%%%%%%%%%%%%%%%%%%%%%%%%%%%%%%%%

\section{A brief review of distribution of the sample coefficient of variation}
\label{sec:CVdistribution}
In this section, the distribution of the CV is briefly  presented. The CV of a random variable $X$, say $\gamma$, is defined as the ratio of the standard deviation $\sigma=\sigma(X)$ to the mean $\mu=E(X)$; i.e.,
\[
\gamma=\frac{\sigma}{\mu}.
\]

Suppose that a sample of size $n$ of normal
i.i.d. random variables $\{X_1,\ldots,X_n\}$ is collected. 
Let $\bar{X}$ and $S$ be the sample mean and the sample 
standard deviation of these variables, i.e.,
\[
\bar{X}=\frac{1}{n}\sum_{i=1}^n X_i
\]

and
\[
S=\sqrt{\frac{1}{n-1}\sum_{i=1}^n(X_i-\bar{X})^2}.
\]

Then the sample coefficient of variation $\hat{\gamma}$ of these variables
is defined as
\[
\hat{\gamma}=\frac{S}{\bar{X}}.
\]

The probability distribution of the sample CV $\hat{\gamma}$ has been studied in the literature 
by many authors. However, the exact distribution of $\hat{\gamma}$ has a complicated
form. The approximate distribution is then widely used as an alternative. In this study, we apply the following approximation of  $F_{\hat{\gamma}}(x|n,\gamma)$, the $c.d.f$ (cumulative distribution function) of  $\hat{\gamma}$, which is suggested by \cite{Castagliola_EWMA_CV_2011}:
 
\begin{equation}
\label{equ:cdfcv}
F_{\hat{\gamma}}(x|n,\gamma)=1-F_t\left(\left.\frac{\sqrt{n}}{x}\right|n-1,
\frac{\sqrt{n}}{\gamma}\right),
\end{equation}
\noindent
where  $F_t\left(.
\left|n-1,\frac{\sqrt{n}}{\gamma}\right.\right)$ is the c.d.f. of the noncentral $t$ distribution 
with $n-1$ degrees of freedom and noncentrality parameter. 
This approximation is only sufficiently precise when $\gamma<0.5$. This condition is in general satisfied in our case as it is very frequent that the CV takes small values to ensure the stability of a process. More details on this problem have been discussed in \cite{Tran_2018_CVME}.
 % $\frac{\sqrt{n}}{\gamma}$. % With some manipulations, inverting $F_{\hat{\gamma}}(x|n,\gamma)$ gives the inverse c.d.f. $F^{-1}_{\hat{\gamma}}(\alpha|n,\gamma)$ of $\hat{\gamma}$ as (\citet{castagliola2011}) 
%  \begin{equation}
%\label{equ:idfcv}
%F^{-1}_{\hat{\gamma}}(\alpha|n,\gamma)=\frac{\sqrt{n}}{F^{-1}_t\left(1-\alpha
%\left|n-1,\frac{\sqrt{n}}{\gamma}\right.\right)},
%\end{equation}
%\noindent
%where $F^{-1}_t\left(.
%\left|n-1,\frac{\sqrt{n}}{\gamma}\right.\right)$ is the inverse c.d.f. of the noncentral $t$
%distribution.\\
  
For the case of the sample CV squared ($\hat{\gamma}^2$), Castagliola et al. \cite{Castagliola_EWMA_CV_2011} showed that $\frac{n}{\hat{\gamma}^2}$ follows a noncentral $F$
distribution with $(1,n-1)$ degrees of freedom and noncentrality
parameter $\frac{n}{\gamma^2}$. Then, they deduced the
$c.d.f$ $F_{\hat{\gamma}^2}(x|n,\gamma)$ of $\hat{\gamma}^2$ as

\begin{equation}
\label{equ:cdfcv2}
F_{\hat{\gamma}^2}(x|n,\gamma)=1-F_F\left(\left.\frac{n}{x}\right|1,n-1,
\frac{n}{\gamma^2}\right),
\end{equation}
\noindent
where $F_F\left(.
\left|1,n-1,\frac{n}{\gamma^2}\right.\right)$ is the $c.d.f$ of the noncentral $F$ distribution. The corresponding density function of  $\hat{\gamma}^2$ is then
\begin{equation}
f_{\hat{\gamma}^2}(x|n,\gamma)=\frac{n}{x^2} f_F\left(\left.\frac{n}{x}\right|1,n-1,
\frac{n}{\gamma^2}\right),
\end{equation}
\noindent
where $f_F\left(.
\left|1,n-1,\frac{n}{\gamma^2}\right.\right)$ is the density function of the noncentral $F$ distribution

\begin{figure}
\centering
\includegraphics[width=\linewidth]{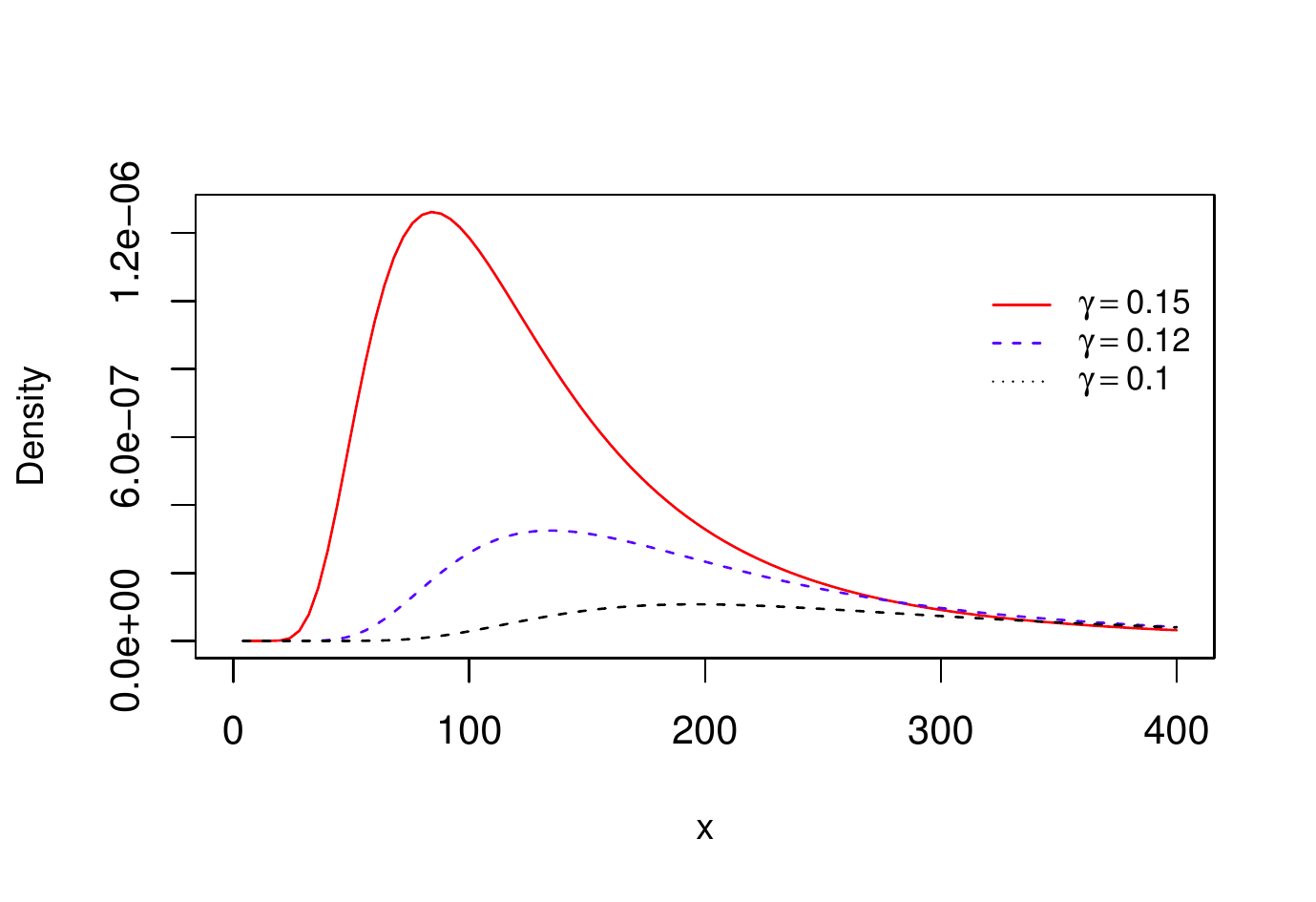}
\caption{The approximate density function of the $\hat{\gamma}^2$ for $n=5$.\label{fig: densityCV2}}
\end{figure}

Figure \ref{fig: densityCV2} presents the density distribution of $\gamma^2$ for $n=5$ and some different values of $\gamma_0$.\\

PLEASE INSERT FIGURE  \ref{fig: densityCV2} HERE
% \citet{castagliola2011} showed that the inverse c.d.f. 
%$F^{-1}_{\hat{\gamma}^2}(\alpha|n,\gamma)$ of $\hat{\gamma}^2$ can be obtained as
%\begin{equation}
%\label{equ:idfcv2}
%F^{-1}_{\hat{\gamma}^2}(\alpha|n,\gamma)=\frac{n}{F^{-1}_F\left(1-\alpha
%\left|1,n-1,\frac{n}{\gamma^2}\right.\right)},
%\end{equation}
%\noindent
%where $F^{-1}_F\left(.
%\left|1,n-1,\frac{n}{\gamma^2}\right.\right)$ is the inverse c.d.f. of the noncentral $F$
%distribution.
%%%%%%%%%%%%%%%%%%%%%%%%%%%%%%%%%%%%%
%%%%%%%%%%%%%%%%%%%%%%%%%%%%%%%%%%%

\section{Design and implementation of the RR$_{r,s}-{\gamma}^2$ control chart}
\label{sec: design no ME}
In the literature, the Run Rules control charts monitoring the CV has been investigated in \cite{Castagliola_Run-Rules_CV_2013} with two-sided charts. However, since the distribution of $\gamma^2$ is asymmetric (as can be seen from the equation (\ref{equ:cdfcv2}) and also from  Figure \ref{fig: densityCV2}), these two-sided charts lead to the problem of $ARL$-biased (Average Run Length) performance, i.e. the out-of-control $ARL_1$ values are larger than the in-control values $ARL_0$. This problem was also pointed out in \cite{Castagliola_Run-Rules_CV_2013}. Therefore, we overcome this $ARL$-biased property by designing simultaneously two one-sided charts to detect both the increase and the decrease at the CV squared. In particular,  we suggest defining two one-sided Run Rules control charts monitoring the CV squared, involving:
\begin{itemize}
\item a lower-sided $r$-out-of-$s$ Run Rules control chart (denoted as
  RR$_{r,s}^--{\gamma}^2$) to detect a decrease in $\hat{\gamma}_i$ with a
  lower control limit $LCL^-=\mu_0(\hat{\gamma}^2)-k_d.\sigma_0(\hat{\gamma}^2)$ and a corresponding upper control
  limit $UCL^-=+\infty$,
\item an upper-sided $r$-out-of-$s$ Run Rules control chart (denoted as
  RR$_{r,s}^+-{\gamma}^2$) to detect a decrease in $\hat{\gamma}_i$ with a
  lower control limit $UCL^+=\mu_0(\hat{\gamma}^2)+k_u.\sigma_0(\hat{\gamma}^2)$ and a corresponding lower control
  limit $LCL^+=-\infty$,
\end{itemize}
where $k_d>0$  and $k_u>0$ are the chart parameters of the RR$_{r,s}^--{\gamma}^2$
and RR$_{r,s}^+-{\gamma}^2$ charts, respectively.

The closed forms of $\mu_0(\hat{\gamma}^2)$ and
$\sigma_0(\hat{\gamma}^2)$ have not been presented in the literature. We apply in this study the accurate approximations provided by Breunig \cite{breunig2001} for both $\mu_0(\hat{\gamma}^2)$ and
$\sigma_0(\hat{\gamma}^2)$  as follows:
\begin{eqnarray}
\label{equ:mu0}
\mu_0(\hat{\gamma}^{2}) & =  &\gamma_0^{2}\bigg(1-\frac{3\gamma_0^{2}}{n}\bigg),\\
\sigma_0(\hat{\gamma}^{2})  &= & \sqrt{\gamma_0^{4}\left(\frac{2}{n-1}+\gamma_0^{2}
\left(\frac{4}{n}+\frac{20}{n(n-1)}+\frac{75\gamma_0^{2}}{n^2}\right)
\right)-(\mu_0(\hat{\gamma}^{2})-\gamma_0^{2})^2}.
\label{equ:sigma0}
\end{eqnarray}

Given the value of the control limit for each chart, an out-of-control signal is given at time $i$ if $r$-out-of-$s$ consecutive $\hat{\gamma}_i$  values are plotted outside the control interval, i.e. $\hat{\gamma}_i^2<LCL^-$ in the lower-sided chart and $\hat{\gamma}_i^2>UCL^+$ in the upper-sided chart. The control charts designed above is called $pure$ Run Rules type chart. In this study, we only consider the 2-out-of-3, 3-out-of-4 and 4-out-of-5 Run Rules charts.  The performance of the proposed charts is measured by the $ARL$ which is calculated by using Markov chain as follows.

Firstly, we define the matrix  $\mathbf{P}$ of the embedded Markov chain. For the two one-sided RR$_{2,3}-{\gamma}^2$ control charts,   $\mathbf{P}$ is defined by \begin{equation}
\label{equ:P}
\mathbf{P}=\left(
  \begin{array}{cc}
  \mathbf{Q} & \mathbf{r} \\
  & \\
  \mathbf{0}^T & 1
  \end{array}
\right)
=\left(
  \begin{array}{ccc|c}
  0 & 0 & p & 1-p \\
  p & 0 & 0 & 1-p \\
  0 & 1-p & p & 0 \\
  \hline
  0 & 0 & 0 & 1
  \end{array}
\right),
\end{equation}
where $\mathbf{Q}$ is a $(3,3)$ matrix of
transient probabilities, $\mathbf{r}$ is a $(3,1)$ vector satisfied
$\mathbf{r}=\mathbf{1}-\mathbf{Q}\mathbf{1}$ with $\mathbf{1}=(1,1,1)^T$ and $\mathbf{0}=(0,0,0)^T$, $p$ is the probability that a sample drops into the control interval. The corresponding $(3,1)$ vector $\mathbf{q}$ of initial probabilities associated with the transient states is  $\mathbf{q}=(0,0,1)^T$, i.e. the third state is the initial state. 

For the case of RR$_{3,4}-{\gamma}^2$ control charts, the transient probability matrix $\mathbf{Q}_{(7\times 7)}$ is given by
\begin{equation}
\mathbf{Q}=\left(
\begin{array}{ccccccc}
0 & 0 & p & 0 & 0 & 0 & 0 \\
0 & 0 & 0 & 0 & p & 0 & 0 \\
0 & 0 & 0 & 0 & 0 & 1-p & p \\
p & 0 & 0 & 0 & 0 & 0 & 0 \\
0 & 1-p & p & 0 & 0 & 0 & 0 \\
0 & 0 & 0 & 1-p & p & 0 & 0 \\
0 & 0 & 0 & 0 & 0 & 1-p & p \\
\end{array}
\right).
\end{equation}
In this case, the seventh state in the vector $\mathbf{q}=(0,0,0,0,0,0,1)^T$  is the initial state.

Extended to ``longer" $(4,5)$ Run Rules, the $(15,15)$ matrix $\mathbf{Q}$ of  transient probabilities for the two one-sided RR$_{4,5}-\gamma^2$ control charts is

\begin{equation}
\label{equ:P}
\small
\mathbf{Q}
=\left(
  \begin{array}{ccccccccccccccc}
1-p & p & 0 & 0 & 0 & 0 & 0&p & 0 & 0 & 0 & 0 & 0 &0&0 \\
0 & 0 & p & 1-p & 0 & 0 & 0&0 & 0 & 0 & 0& 0 & 0 & 0&0 \\
0 & 0 & 0 & 0 &  p & 1-p & 0&0 & 0 & 0 & 0 & 0 & 0 & 0&0 \\
0 & 0 & 0 & 0 & 0 & 0 & 1-p&p & 0 & 0 & 0 & 0 & 0 & 0&0\\
0 & 0 & 0 & 0 & 0 & 0 &0&0 & 0 & 0 & 0 & 0 & 0 & 0&1-p\\
0 & 0 & 0 & 0 &0& 0 & 0&0 & 0 & 0 & 0 & 0 & p & 1-p&0\\
0 & 0 & 0 & 0 & 0 & 0 & 0&0 & 0 & 0 & 0 &p& 1-p & 0&0 \\
0 &0& $0$ & 0 & 0 & 0 & 0&1-p & p & 0 & 0 & 0 & 0 & 0&0\\
0 & 0 & 0 & 0 & 0 & 0 & 1-p&p & 0 &0& 0 & 0 & 0 & 0&0 \\
0 & 0 & 0 & 0 & p & 1-p & 0&0 & 0 & 0 & 0 & 0 &0& 0&0 \\
0 & 0 & p & 1-p & 0 & 0 & 0&0 & 0 & 0 & 0 &0& 0 & 0&0 \\
1-p & p & 0 & 0 & 0 & 0 & 0&0 & 0 & 0 & 0 & 0 & 0 & 0&0 \\
0 & 0 & 0 & 0 &0& 0 & 0&0 & 0 & 1-p & p & 0 & 0 & 0&0 \\
0 & 0 & 0 & 0 & 0 & 0 & 0&0 & 0 & 0 & 0 & p & 1-p & 0&0 \\
0 & 0 & 0 & 0 & 0 & 0 & 0&0 &0& 0 & 0 & 0 & 0 & p&1-p \\
  \end{array}
\right),
\end{equation}
that corresponds to the $(15,1)$ initial probabilites vector $\mathbf{q}=(0,\ldots,0,1)^T$ (i.e. the initial
state is the 15th one). These transient probability matrices has been presented in, for example, \cite{Tran2016_Runrules_RZ}, \cite{Tran2017Runrules_median_mean}, \cite{Tran_2018_runrule_t}.

Let us now suppose that the occurrence of an unexpected condition shifts the in-control CV value $\gamma_0$  to the out-of-control value $\gamma_1=\tau\times \gamma_0$, where $\tau>0$ is the shift size. Values of $\tau\in (0,1)$ correspond to a decrease of the $\gamma_0$, while values of $\tau>1$ correspond to an increase of $\gamma_0$. Then, the probability $p$ is defined by
\begin{itemize}
\item for the  RR$_{r,s}^--{\gamma}^2$ chart:
  \begin{equation} \label{p1}
  p=P(\hat{\gamma}_i^2\geq LCL^-)= 1- F_{\hat{\gamma}^2}(LCL^-|n,\gamma_1),
  \end{equation}
\item for the  RR$_{r,s}^+-{\gamma}^2$ chart:
  \begin{equation} \label{p2}
 p=P(\hat{\gamma}_i^2\leq UCL^+)= F_{\hat{\gamma}^2}(UCL^+|n,\gamma_1),
  \end{equation}
\end{itemize}
where  $F_{\hat{\gamma}^2}$ is defined in  (\ref{equ:cdfcv2}).

Once the matrix $\mathbf{Q}$ and the vector $\mathbf{q}$ have been determined, the $ARL$ and the $SDRL$ (standard deviation of run length) are calculated by
\begin{eqnarray} 
\label{ARL}
ARL & = & \mathbf{q}^T(\mathbf{I}-\mathbf{Q})^{-1}\mathbf{1}, \\
SDRL & = & \sqrt{2\mathbf{q}^T(\mathbf{I}-\mathbf{Q})^{-2}\mathbf{Q}\mathbf{1}-ARL^2+ARL}.
\end{eqnarray}

%%with
%\begin{eqnarray}
%\nu_1 & = & \mathbf{q}^T(\mathbf{I}-\mathbf{Q})^{-1}\mathbf{1},\\
%\nu_2 & = & 2\mathbf{q}^T(\mathbf{I}-\mathbf{Q})^{-2}\mathbf{Q}\mathbf{1},\\
%\mu_2 & = & \nu_2-\nu_1^2+\nu_1.
%\end{eqnarray}

It is customary that a control chart is considered to be better than its competitors if it gives a smaller value of the $ARL_1$ while the $ARL_0$ is the same. Therefore, the parameters of the RR$_{r,s}-{\gamma}^2$ control charts should be the solution of the following equations:
\begin{itemize}
\item for the  RR$_{r,s}^--{\gamma}^2$ chart:
  \begin{equation} \label{equ:control limit1}
  ARL(k_d,n,p,\gamma_0,\tau=1)=ARL_0,
  \end{equation}
\item for the  RR$_{r,s}^+-{\gamma}^2$ chart:
  \begin{equation} \label{equ:control limit2}
 ARL(k_u,n,p,\gamma_0,\tau=1)=ARL_0,
  \end{equation}
\end{itemize}
where $ARL_0$ is predefined.

%%%%%%%%%%%%%%%%%%%%%%%%%%%%%%%%%%%%%%%%%%%%%%%%%%%%%%%%%%%%%%%%%%%%%%%%%%%%%%%%
\section{Performance of RR$_{r,s}-{\gamma}^2$ control charts}
\label{sec: perform no ME}
 %-----------------------------------------  
Assigning the in-control value $ARL_0$ at $ARL_0=370.4$,  the parameters $k_d$ and $k_u$ of the lower-sided and upper-sided RR$_{r,s}-{\gamma}^2$ charts for some combinations of $n \in \{5,15\}, \gamma_0 \in \{0.05,0.1,0.2\}$ are presented in Table \ref{tab:nome}. Table \ref{tab:ARL1} shows the corresponding $ARL_1$ values of the proposed charts for various situations of the shift size $\tau$. The obtained results show that the two one-sided RR$_{r,s}-{\gamma}^2$ charts not only overcome the $ARL$-biased problem (as the $ARL_1$ values are always smaller than the $ARL_0$) but also outperform the two-sided RR-$\gamma$ charts investigated in \cite{Castagliola_Run-Rules_CV_2013}. For example, with $\gamma_0=0.05, \tau=1.10$ and $n=5$ in the RR$_{2,3}-{\gamma}^2$ chart, we have $ARL_1=95.9$ ( Table \ref{tab:ARL1} in this study), which is smaller than $ARL_1=101.6$ (Table 2 in \cite{Castagliola_Run-Rules_CV_2013}).\\
%%%Table1%%%
\begin{center}
  INSERT TABLE \ref{tab:LUCL} ABOUT HERE
\end{center}
%%%Table2%%%
\begin{center}
  INSERT TABLE \ref{tab:nome} ABOUT HERE
\end{center}
%%%Table3%%%
\begin{center}
  INSERT TABLE \ref{tab:ARL1} ABOUT HERE
\end{center}
%%%%%%%%%%%%%%%%%%%%%%%%%%%%%%%%%%%%%%%%%%%%
%%%%%%%%%%%%%%%%%%%%%%%%%%%%%%%%%%%%%%%%%%%%%%%%%%%%%%%%%%%%%%%%%%%%%%%%%%%%%
%%%%%%%%%%%%%%%%%%%%%%%%%%%%%%%%%%%%%%%%%%%%%%%%%%%%%%%%%%%%%%%%%%%%%%%%%%%%%
\section{Linear covariate error model for the coefficient of variation}
\label{sec: ME model}
The previous design for the RR$_{r,s}-{\gamma}^2$ control charts is based on a latent assumption that the values in the collected sample are measured exactly without the measurement error. This assumption, however, is usually not reached in practice and it is difficult to avoid the measurement error. That leads to many authors have conducted their studies based on the measurement error presence, see for instance \cite{Nguyen_2019_VSICV_ME, tran2019synthetic}

In this section, we suppose a linear covariate error model to the measurement error, which is suggested by \cite{linna2001effect}.

Suppose that the quality characteristic $X$ of $n$ consecutive items at step i$^{th}$ is $(X_{i,1}, X_{i,2},.., X_{i,n})$, where $X_{i,j}\sim N(\mu_0+a\sigma_0,b^2\sigma_0^2)$ where $\mu_0$ and $\sigma_0$ are the in-control mean and standard deviation of $X$ and they are independent. The parameters $a$ and $b$ represent the standardized mean and standard deviation shift. The process has shifted if the process mean $\mu_0$ 
and/or the process standard deviation $\sigma_0$ have changed ($a\ne 0$ and/or $b\ne 1$). Due to the measurement error, we only observe the values $(X_{i,j,1}^*,...,X_{i,j,m}^*)$ of a set of $m$ measurement operations instead the true values $X_{i,j}$. According to the linear covariate error model, we assume $X_{i,j,k}^* = A + BX_{i,j} + \varepsilon_{i,j,k}$,
where $A$ and $B$ are two known constants and $\varepsilon_{i,j,k}$ is a normal random error term with parameters $(0, \Sigma_M)$ and independent of $X_{i,j}$. 

Let $\bar{X}_{i,j}^*$ denote the mean of $m$ observed quantities of the same item $j$ at the i$^{th}$ sampling. It is straightforward to show that
\begin{equation*}
\bar{X}_{i,j}^*  ~\sim ~ N(\mu^*,\sigma^{*2}) =  N\left(A+B(\mu_0+a\sigma_0),B^2b^2\sigma^2_0+\frac{\sigma^2_M}{m}\right).
\end{equation*}
\cite{Tran_2018_CVME} showed that the CV of the quantity $\bar{X}_{i,j}^*$ is
\begin{equation}
\label{equ:CV ME}
\gamma^*=\frac{\sigma^* }{\mu^*}= \frac{\sqrt{B^2b^2+\frac{\eta^2}{m}}}{\theta+B(1+a\gamma_0)}
  \times\gamma_0.
\end{equation} 
where $\gamma_0=\frac{\sigma_0}{\mu_0},\eta=\frac{\sigma_M}{\sigma_0}$ and $\theta=\frac{A}{\mu_0}$ are the in-control value of CV, the precision error ratio and the accuracy error, respectively.
The sample coefficient of variation $\hat{\gamma}^*_i$
is defined by $\hat{\gamma}^*_i=\frac{S_i^*}{\bar{\bar{X}}_i^*}$ where $\bar{\bar{X}}_i^*$ and $S_i^*$
%in which 
%\[
%\bar{\bar{X}}_i^*=\frac{1}{n}\sum_{j=1}^n \bar{X}_{i,j}^*
%~~\text{and}~~
%S_i^*=\sqrt{\frac{1}{n-1}\sum_{j=1}^n(\bar{X}_{i,j}^*-\bar{\bar{X}}_i^*)^2}
%\]
%\noindent
are the sample mean and the sample standard deviation of $\bar{X}_{1,j}^*,\ldots,\bar{X}_{n,j}^*$.
The  $c.d.f$ of $\hat{\gamma}^{*2}$ can be obtained from
(\ref{equ:cdfcv2}) by simply
replacing $\gamma$ by $\gamma^*$, i.e., the $c.d.f$ $F_{\hat{\gamma}^{*2}}(x|n,\gamma^*)$  of $\hat{\gamma}^{*2}$ is given by

\begin{equation}
\label{equ:cdfcv2me}
F_{\hat{\gamma}^{*2}}(x|n,\gamma^*)=1-F_F\left(\frac{n}{x}
\left|1,n-1,\frac{n}{\gamma^{*2}}\right.\right)
\end{equation}

%%%%%%%%%%%%%%%%%%%%%%%%%%%%%%%%%%%
\section{Implementation and the performance of the RR$_{r,s}-{\gamma}^2$ charts with measurement errors}
\label{sec:perform me}
Under the
presence of measurement errors, the values $\mu_0(\hat{\gamma}^{*2})$ and $\sigma_0(\hat{\gamma}^{*2})$ are calculated as in  (\ref{equ:mu0}) and (\ref{equ:sigma0}), where $\gamma_0$ is replaced by $\gamma_0^*$, which is defined from \eqref{equ:CV ME} with $a=0$ and $b=1$:
\begin{equation}
\gamma^*_0= \frac{\sqrt{B^2+\frac{\eta^2}{m}}}{\theta+B)}
  \times\gamma_0.
\end{equation} 

Suppose that the in-control value $\gamma_0$ is shifted to the out-of-control value $\gamma_1$ with the size $\tau$, we can represent $\tau$ according to $a$ and $b$ as $\tau=b/(1+a\gamma_0)$. Therefore, the out-of-control CV of the observed quantity $\bar{X}^*_{i,j}$ can be expressed by
\begin{equation}
\gamma^*_1= \frac{\sqrt{B^2b^2+\frac{\eta^2}{m}}}{\theta+\frac{Bb}{\tau}}
  \times\gamma_0.
\end{equation} 
In the implementation of RR$_{r,s}-{\gamma}^2$ control charts, the control limits,  $UCL^{*+}=\mu_0(\hat{\gamma}^{*2})+k_u^*.\sigma_0(\hat{\gamma}^{*2})$ and $LCL^{*-}=\mu_0(\hat{\gamma}^{*2})-k_d^*.\sigma_0(\hat{\gamma}^{*2})$, are also found by solving the chart parameters $k_d$ and $k_u$ as the solution of the following equations 
\begin{itemize}
\item for the  RR$_{r,s}^--{\gamma}^2$ chart:
  \begin{equation} \label{equ:control limit11}
  ARL(k_d,n,p,\gamma_0,\theta,\eta,m,B,b)=ARL_0,
  \end{equation}
\item for the  RR$_{r,s}^{+}-{\gamma}^2$ chart:
  \begin{equation} \label{equ:control limit22}
ARL(k_u,n,p,\gamma_0,\theta,\eta,m,B,b)=ARL_0.
  \end{equation}
\end{itemize}
The $ARL$ in \eqref{equ:control limit11} and  \eqref{equ:control limit22} should be calculated with the transition probability matrix $\mathbf{Q}$ where the transition probability $p$  is defined from (\ref{p1}) and (\ref{p2}) but with the $c.d.f$ $F_{\hat{\gamma}^{*2}}(x|n,\gamma^*)$  of $\hat{\gamma}^{*2}$ in (\ref{equ:cdfcv2me}) instead of $c.d.f$ $F_{\hat{\gamma}^2}$ in  (\ref{equ:cdfcv2}).

To investigate the performance of the RR$_{r,s}-{\gamma}^2$ charts under the appearance of the measurement error, we consider several possible values of the parameters: $n\in \{5,15\}$, $\gamma_0\in \{0.05,0.1,0.2\}$, $\eta\in \{0,0.1,0.2,0.3,0.5,1\}$, $\theta\in \{0,0.01,0.02,0.03,0.04,0.05\}$, $m\in\{1,3,5,7,10\}$ and $B\in\{0.8,0.9,1,1.1,1.2\}$. Without loss of generality, we assume in the remaining that $b=1$. The in-control value CV is also set at $ARL_0=370.4$. 

The control limits of the proposed charts for some specific values of these parameters have been presented in Table \ref{tab:LUCL}. The other values of the control limits for other situations of these parameters are not presented here but are available upon request from authors. 

Tables \ref{tab:eta}-\ref{tab:m} show the corresponding values of the $ARL_1$ under different effects of the parameters $\eta, \theta, m$ and $B$ of the linear covariate model. Some simple conclusions can be drawn from these tables as follows.
\begin{itemize}
\item The increase of the precision error ratio $\eta$ leads to an increase of the $ARL_1$. However, this increase in the $ARL_1$ following the change of $\eta$ is not significant, especially when $\eta\leqslant 0.3$. For example,  for the RR$_{2,3}-\gamma^2$ chart with $n=5, \gamma_0=0.05,B=1,m=1,\theta=0.05$ and $\tau=0.8$, we have $ARL_1=93.12$ when $\eta=0.0$ and $ARL_1=93.20$ when $\eta=0.3$ (Table \ref{tab:eta}). That means the precision error ratio does not affect much the performance of the proposed charts.
\item The accuracy error $\theta$ has a negative impact on the  RR$_{r,s}-\gamma^2$ charts' performance: the larger the accuracy error $\theta$ is, the larger the value $ARL_1$ is, i.e. the lower of the control chart is in detecting the out-of-control condition. For example, in the RR$_{3,4}-\gamma^2$ chart with $n=5, \gamma_0=0.1,B=1,m=1,\eta=0.28$ and $\tau=1.3$, we have $ARL_1=26.56$ when $\theta=0.0$ and $ARL_1=29.19$ when $\theta=0.5$ (Table \ref{tab:theta})
\item Given the value of other parameters, the variation of $B$ significantly affects the performance of the RR$_{r,s}-\gamma^2$ charts. For instance, in Table \ref{tab:B} with the RR$^-_{4,5}-\gamma^2$ control chart and $n=5,m=1,\gamma_0=0.2,\eta=0.28, \theta=0.05, \tau=0.7$ we have $ARL_1=14.43$ when $B=0.8$ and $ARL_1=13.93$ when $B=1.2$.
\item In many situations, taking multiple measurements per item in each sample is an alternative to compensate for the effect of the measurement error. However, the obtained results in this study show that this is not effective way. When $m$ increases from $m=1$ to $m=10$, the $ARL_1$ decreases trivially or is almost unchanged. For example, with $n=5,B=1,\gamma_0=0.05,\eta=0.28, \theta=0.05, \tau=0.8$ in the RR$^+_{2,3}-\gamma^2$ we have $ARL_1=9.07$ for both $m=1$ and $m=10$ (Table \ref{tab:m}).
\item In general, the RR$_{r,s}-\gamma^2$ control charts give better performance in detecting the small process shifts  compared to the VSI-$\gamma^2$ control chart investigated in \cite{Nguyen_2019_VSICV_ME}, under the same condition of measurement errors. For example, with the same values of $n=5,\gamma_0=0.05,\eta=0.28,\theta=0.05,\tau=0.8$, we have $ARL_1=46.80$ for the RR$_{4,5}-\gamma^2$ (Table \ref{tab:theta} in this study), which is smaller than $ARL_1=61.99$ for the VSI $\gamma^2$ control chart with $(h_S,h_L)=0.1,4.0$ (Table 10 in \cite{Nguyen_2019_VSICV_ME}).
\end{itemize}

\begin{center}
  INSERT TABLE \ref{tab:eta} ABOUT HERE
\end{center}
\begin{center}
  INSERT TABLE \ref{tab:theta} ABOUT HERE
\end{center}
\begin{center}
  INSERT TABLE \ref{tab:B} ABOUT HERE
\end{center}
\begin{center}
  INSERT TABLE \ref{tab:m} ABOUT HERE
\end{center}

In practice, quality practitioners often prefer detecting a range of shifts $\Omega = [a; b]$ since it is difficult to guess an exact value for the process shift. In such situations, the statistical
performance of the control chart can be evaluated through the $EARL$ (expected average run length) defined as
\begin{equation}
EARL=\int_{\Omega}ARL\times f_{\tau}(\tau)d\tau,
\end{equation}
where $f_{\tau}(\tau)$ is the distribution of process shift $\tau$ and $ARL$ is defined in (\ref{ARL}). Without any information about $\tau$, one can choose the uniform distribution in $\Omega$, i.e, $f_{\tau}(\tau)=\frac{1}{b-a}$.

\begin{figure*}
\hspace*{-10mm} \includegraphics[scale=0.8]{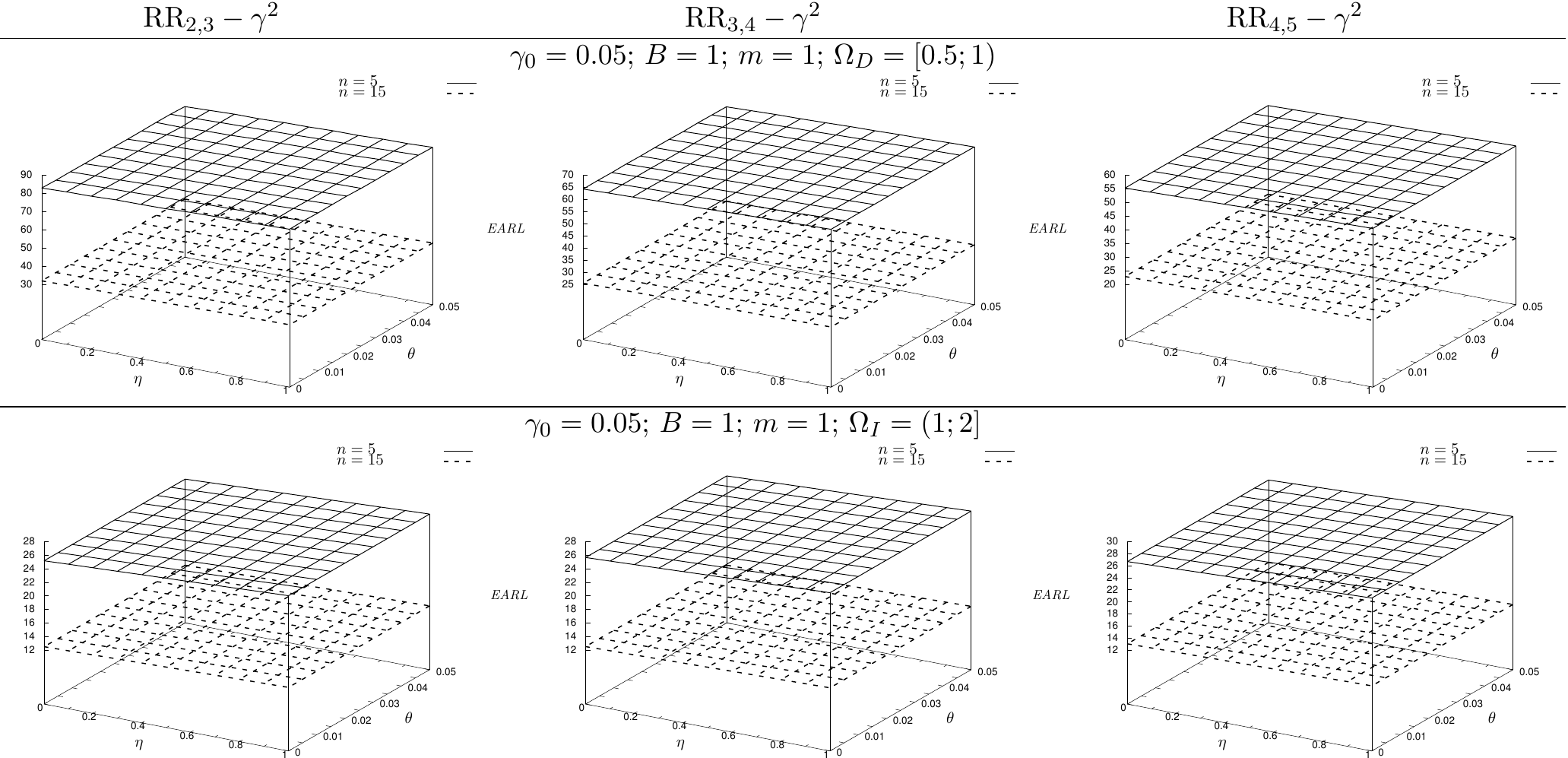}
\caption{The effect of $\theta$ and $\eta$ on the performance of the RR$_{r,s}-\gamma^2$ control charts in the presence of measurement errors for $\gamma_0=0.05$. \label{fig: eta theta 0.05}}
\end{figure*}

\begin{figure*}
\hspace*{-10mm} \includegraphics[scale=0.8]{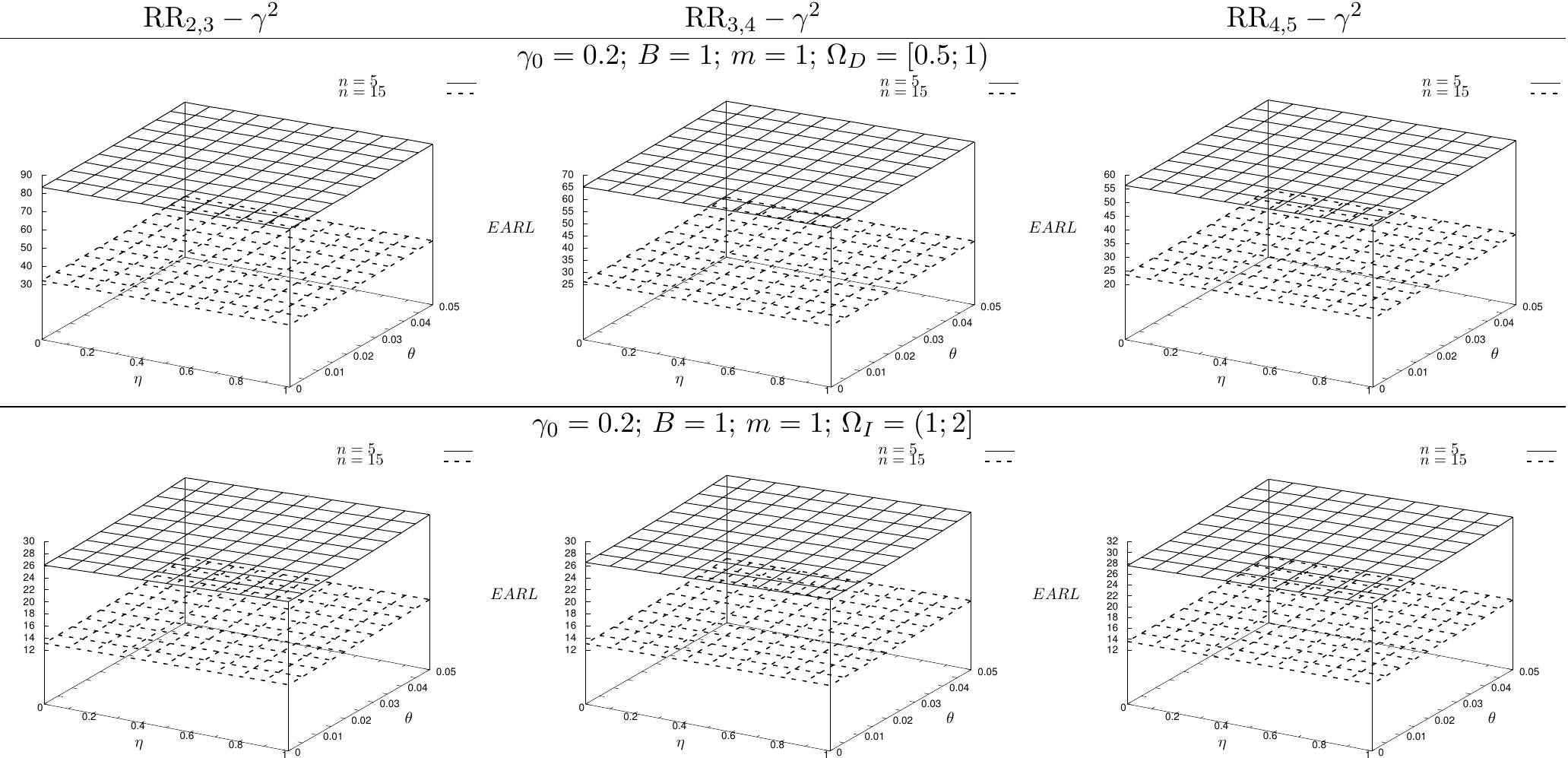}
\caption{The effect of $\theta$ and $\eta$ on the performance of the RR$_{r,s}-\gamma^2$ control charts in the presence of measurement error for $\gamma_0=0.2$.\label{fig: eta theta 0.2}}
\end{figure*}

\begin{figure*}
\hspace*{-10mm}\includegraphics[scale=0.78]{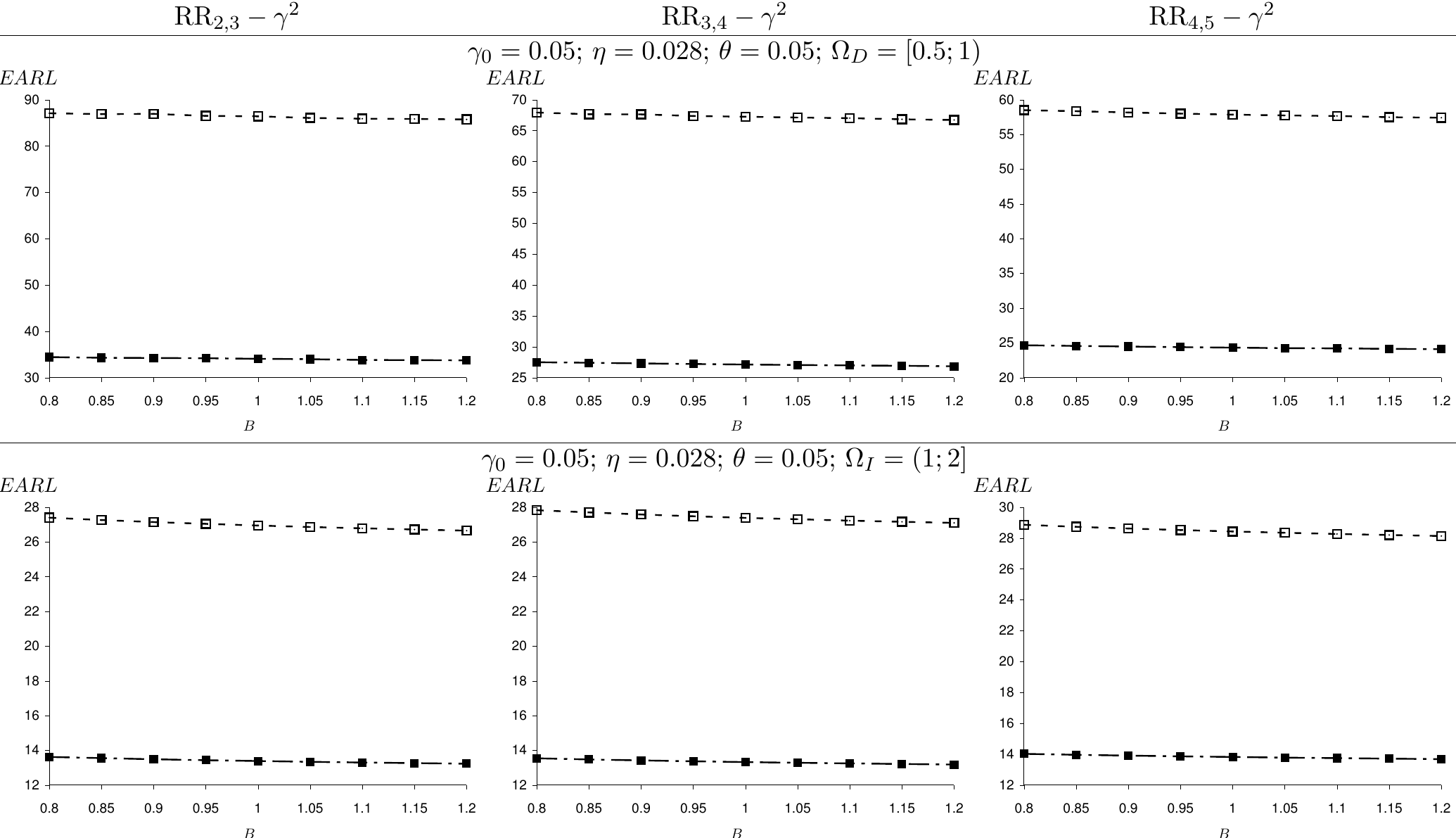}
\caption{The effect of $B$ on the performance of the RR$_{r,s}-\gamma^2$ control charts in the presence of measurement errors for  $\gamma_0=0.05$; $n=5$ (-$\square$-) and $n=15$ ($-\blacksquare-$). \label{fig: B 0.05}}
\end{figure*}

\begin{figure*}
\hspace*{-10mm}\includegraphics[scale=0.78]{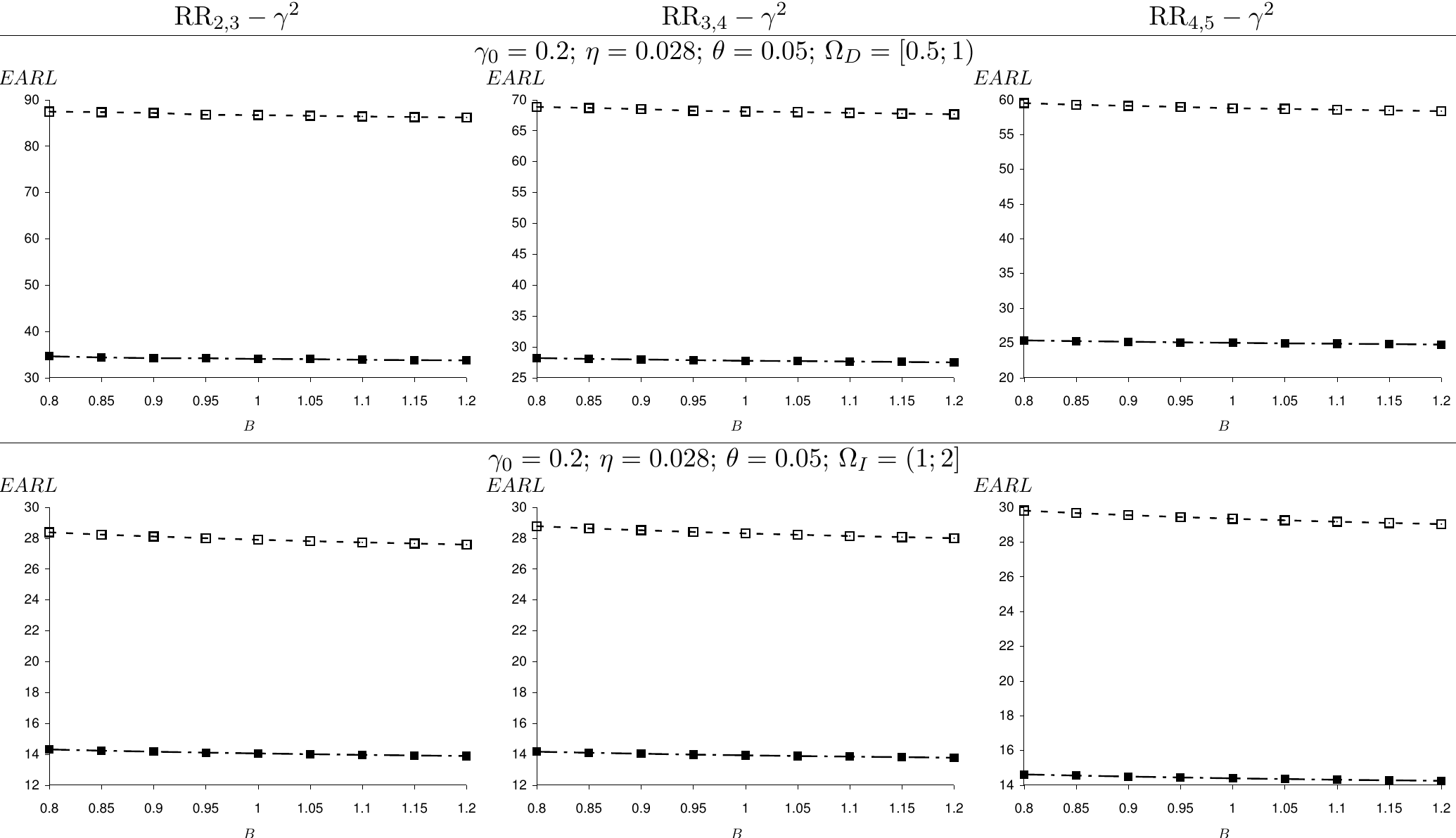}
\caption{The effect of $B$ on the performance of the RR$_{r,s}-\gamma^2$ control charts in the presence of measurement error for  $\gamma_0=0.2$; $n=5$ (-$\square$-) and $n=15$ ($-\blacksquare-$).\label{fig: B 0.2}}
\end{figure*}

\begin{figure*}
\hspace*{-10mm} \includegraphics[scale=0.78]{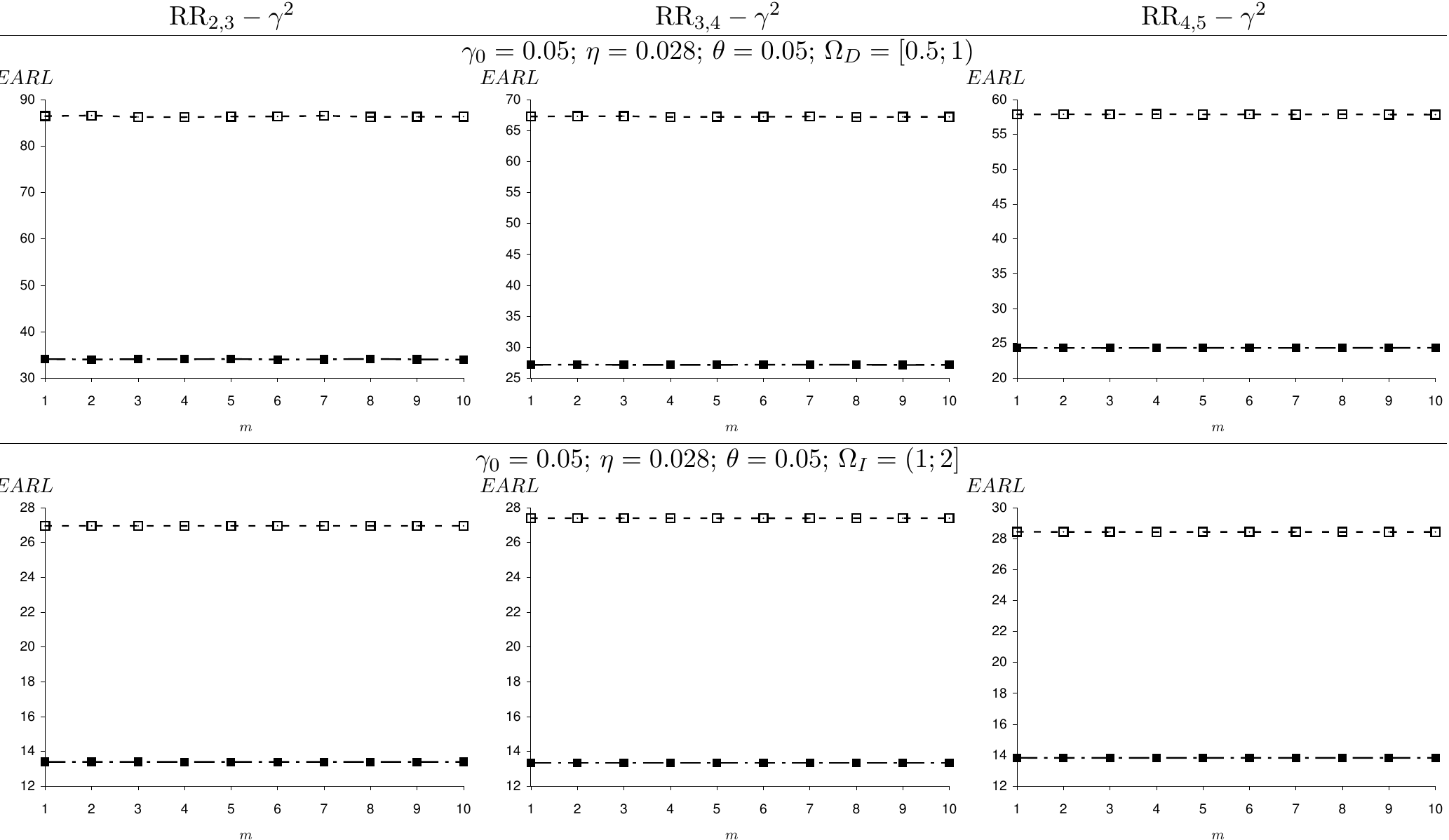}
\caption{The effect of $m$ on the performance of the RR$_{r,s}-\gamma^2$ control charts in the presence of measurement errors for  $\gamma_0=0.05$; $n=5$ (-$\square$-) and $n=15$ ($-\blacksquare-$). \label{fig: m 0.05}}
\end{figure*}

\begin{figure*}
\hspace*{-10mm} \includegraphics[scale=0.78]{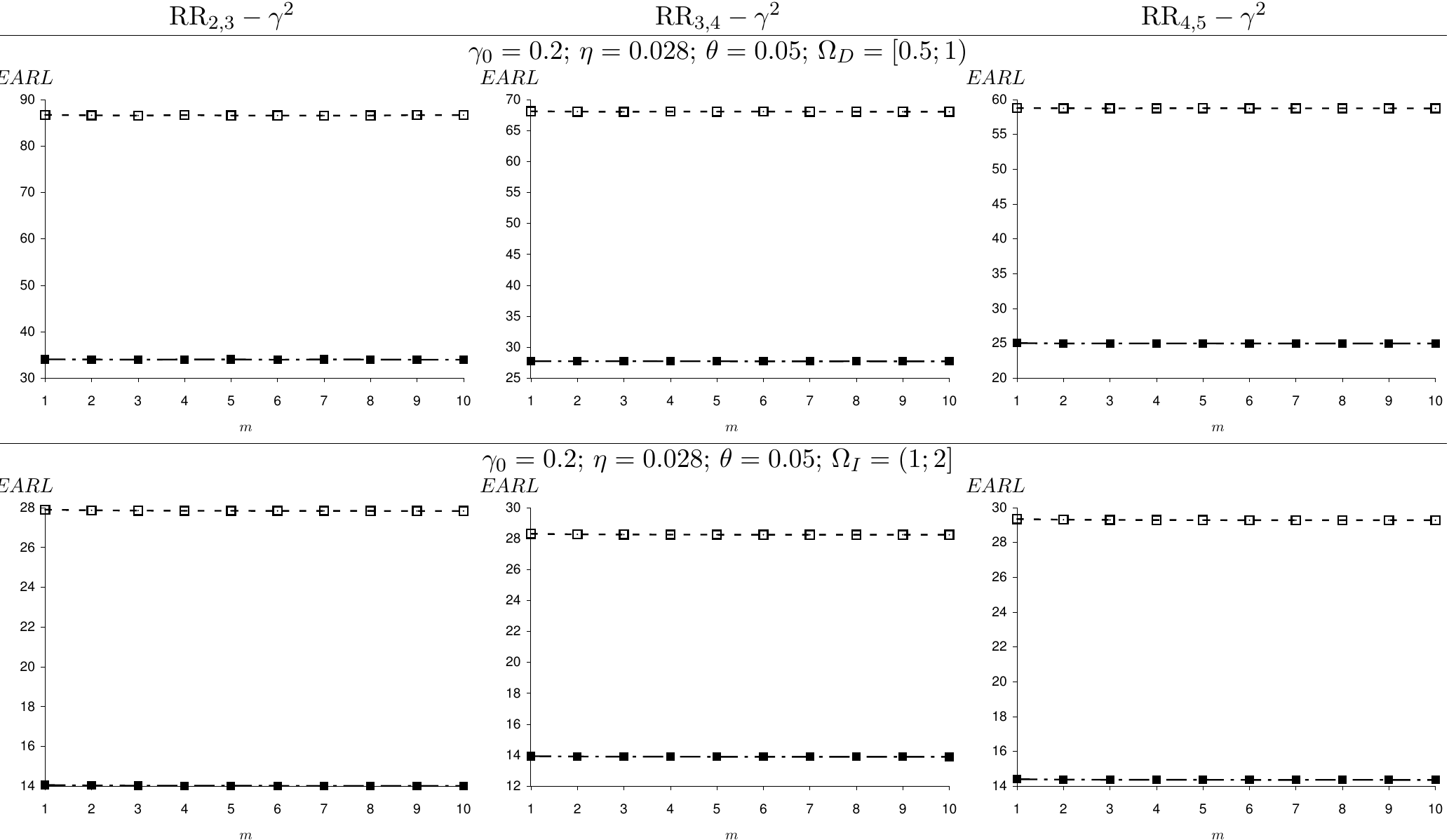}
\caption{The effect of $m$ on the performance of the RR$_{r,s}-\gamma^2$ control charts   in the presence of measurement error for  $\gamma_0=0.2$; $n=5$ (-$\square$-) and $n=15$ ($-\blacksquare-$).\label{fig: m 0.2}}
\end{figure*}

The chart parameters   are  now defined as
\begin{itemize}
\item for the  RR$_{r,s}^--{\gamma}^2$ chart:
  \begin{equation} \label{equ:control limit 1}
  EARL(LCL^{*-},n,p,\gamma_0,\theta,\eta,m,B)=ARL_0,
  \end{equation}
\item for the  RR$_{r,s}^{+}-{\gamma}^2$ chart:
  \begin{equation} \label{equ:control limit 2}
EARL(UCL^{*+},n,p,\gamma_0,\theta,\eta,m,B)=ARL_0.
  \end{equation}
\end{itemize}
In the following simulation, we consider a specific range of decreasing shifts $\Omega_D=[0.5,1)$ and increasing shifts $\Omega_I=(1,2]$. 
Figure \ref{fig: eta theta 0.05}-\ref{fig: eta theta 0.2}  show the change of $EARL$ of the RR-$\gamma^2$ control charts when $\eta$ varies in $[0,1]$ and $\theta$ varies in $[0,0.05]$ for $\gamma_0=0.05$ and $\gamma_0=0.2$, respectively. The slope of the plane which represents the $EARL$ values from right to left and from outside to inside shows that the larger the values of $\eta$ and $\theta$, the larger the value of $EARL$. That is to say, these errors have negative effects on the performance of the RR-$\gamma^2$ charts. For example, in Figure \ref{fig: eta theta 0.05} when $n=5,B=m=1$, and $\gamma=0.05$,  we have $EARL=82.27$ for $\theta=\eta=0$ (corresponding to no measurement errors), but $EARL = 82.81$ for $\eta=0, \theta=0.05$ (corresponding to the negative effect of accuracy error), $EARL = 83.42$ for $\theta=0, \eta=0.3$ (corresponding to the negative effect of precision error), and $EARL = 84.49$ for $\theta=0.05, \eta=0.5$ (corresponding to the negative effect of both precision and accuracy error).
The effect of $B$ and $m$ on the $EARL$  is displayed in Figures \ref{fig: B 0.05}-\ref{fig: m 0.2} for both $\gamma_0=0.05$ and $\gamma_0=0.2$. We obtain a similar trend as the case of the specific shift size: When $B$ increases, the $EARL$ decreases and the $EARL$ does not change significantly when $m$ increases. The almost constant $EARL$ line shows that the effect of $m$ on these chart performance is insignificant. That is to say, increasing the value of $m$ does not reduce the negative effect of measurement errors on the charts. In contrast, the plot of the $EARL$ corresponding to $n=15$ is always below the plot of the $EARL$ corresponding to $n=5$. That means, the sample size has a great impact on the RR$_{r,s}-\gamma^2$ charts' performance regardless of the measurement error.

\begin{center}
PLEASE INSERT FIGURE  \ref{fig: eta theta 0.05} HERE\\
PLEASE INSERT FIGURE  \ref{fig: eta theta 0.2} HERE\\
PLEASE INSERT FIGURE  \ref{fig: B 0.05} HERE\\
PLEASE INSERT FIGURE  \ref{fig: B 0.2} HERE\\
PLEASE INSERT FIGURE  \ref{fig: m 0.05} HERE\\
PLEASE INSERT FIGURE  \ref{fig: m 0.2} HERE
\end{center}

\section{Illustrative example}
\label{sec: example}
%Control charts are widely used in a large number of industrial processes where the out-of-control state should be recognized quickly to guarantee the product quality. In order to monitor these processes, the quantity of interest is calcaluted from the collected data, which commonly contain measurement errors. For the processes where the CV squared is considered, that means only $\gamma^{*2}$ is observable and  the true value $\gamma^2$ is not obtained. However, by monitoring $\gamma^{*2}$, one can still detect changes in the actual CV, and then the process variability. 
  In this section, we present an illustrative example of the implementation of the RR$_{r,s}-\gamma^2$ control charts in the presence of the measurement error. The real industrial data from a sintering process in an Italian company that manufactures sintered mechanical parts, which were introduced in \cite{Castagliola_EWMA_CV_2011}, are considered. 

The process manufactures parts  guarantee a pressure test by  dropping time $T_{pd}$ from 2 bar to 1.5 bar larger than 30 sec as a quality characteristic related to the pore shrinkage. Since the presence of a constant proportionality $\sigma_{pd}=\gamma_{pd}\times\mu_{pd}$ between the standard deviation of the pressure drop time and its mean had been demonstrated  by the preliminary regression study relating $T_{pd}$ to the quantity $Q_C$ of molten copper, the quality practitioners decide to monitor the CV $\gamma_{pd}=\sigma_{pd}/\mu_{pd}$ to detect changes in the process variability. According to the description in \cite{Castagliola_EWMA_CV_2011},
an estimate $\hat{\gamma}_0=0.417$ is calculated from a Phase I dataset based on a root mean square computation. The phase II data are reproduced in Table \ref{tab:data}.

\begin{figure}
  \begin{center}
  \hspace*{-5mm}
    \includegraphics[width=110mm]{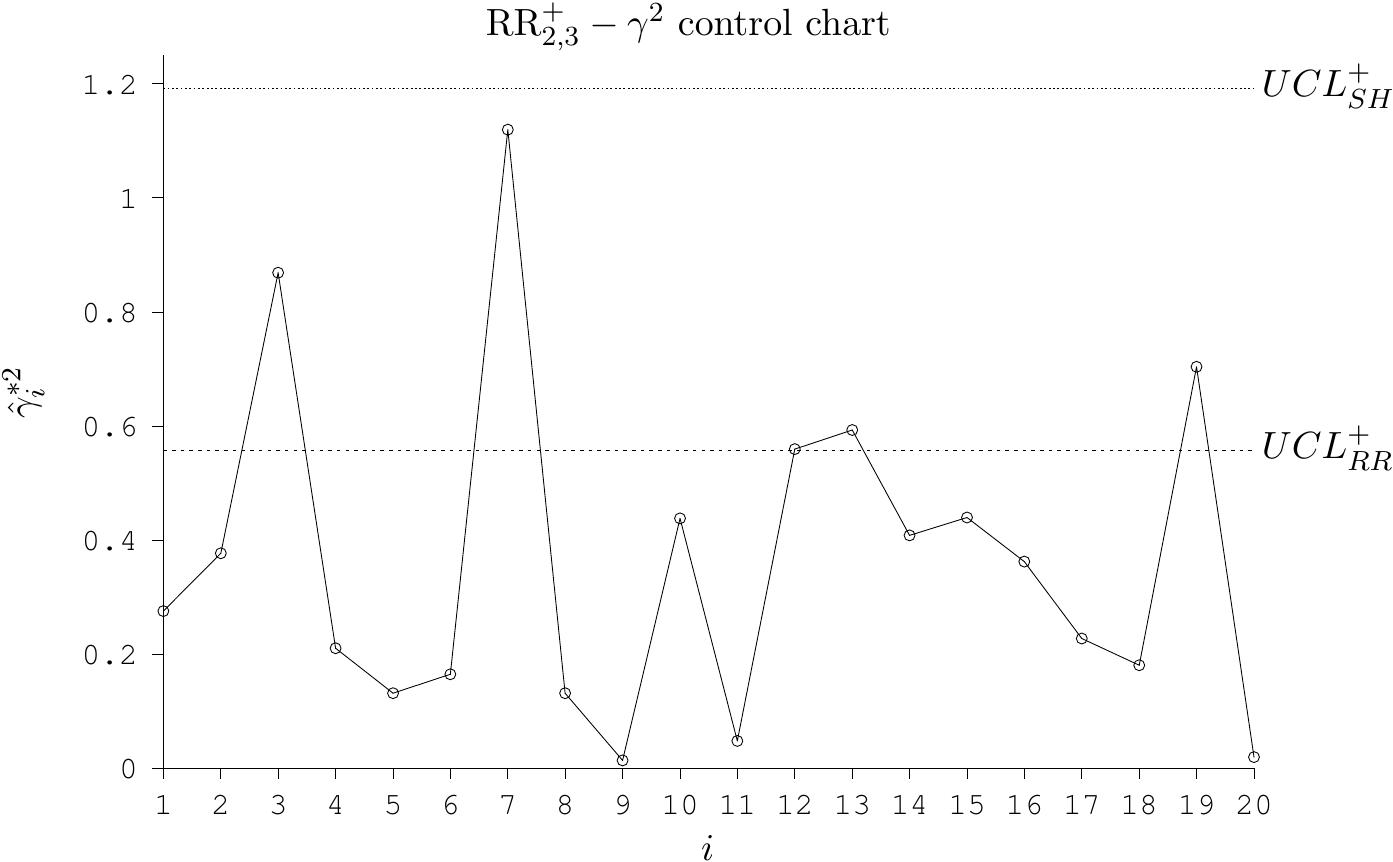} \\[5mm]
     \includegraphics[width=110mm]{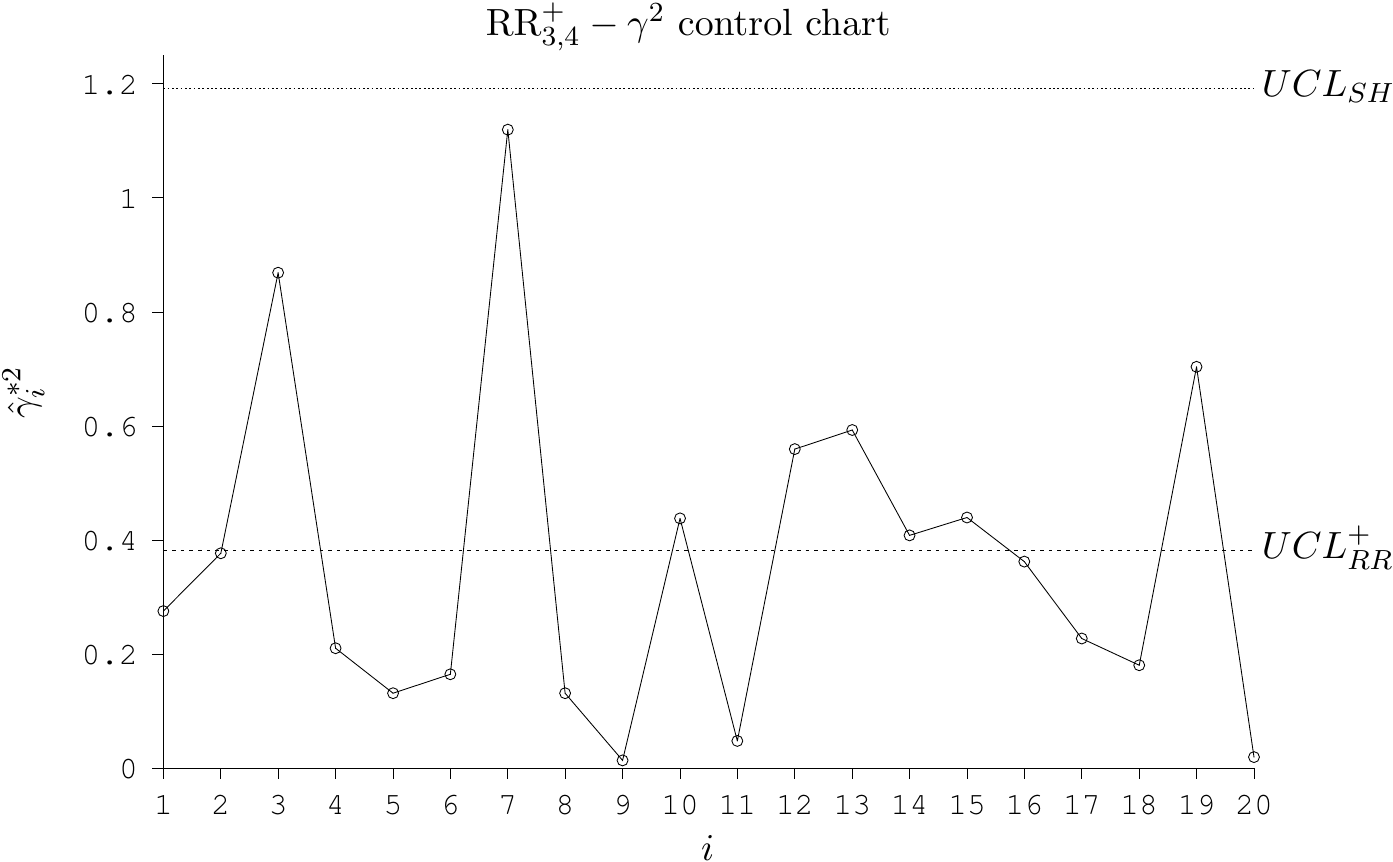} \\[5mm]
      \includegraphics[width=110mm]{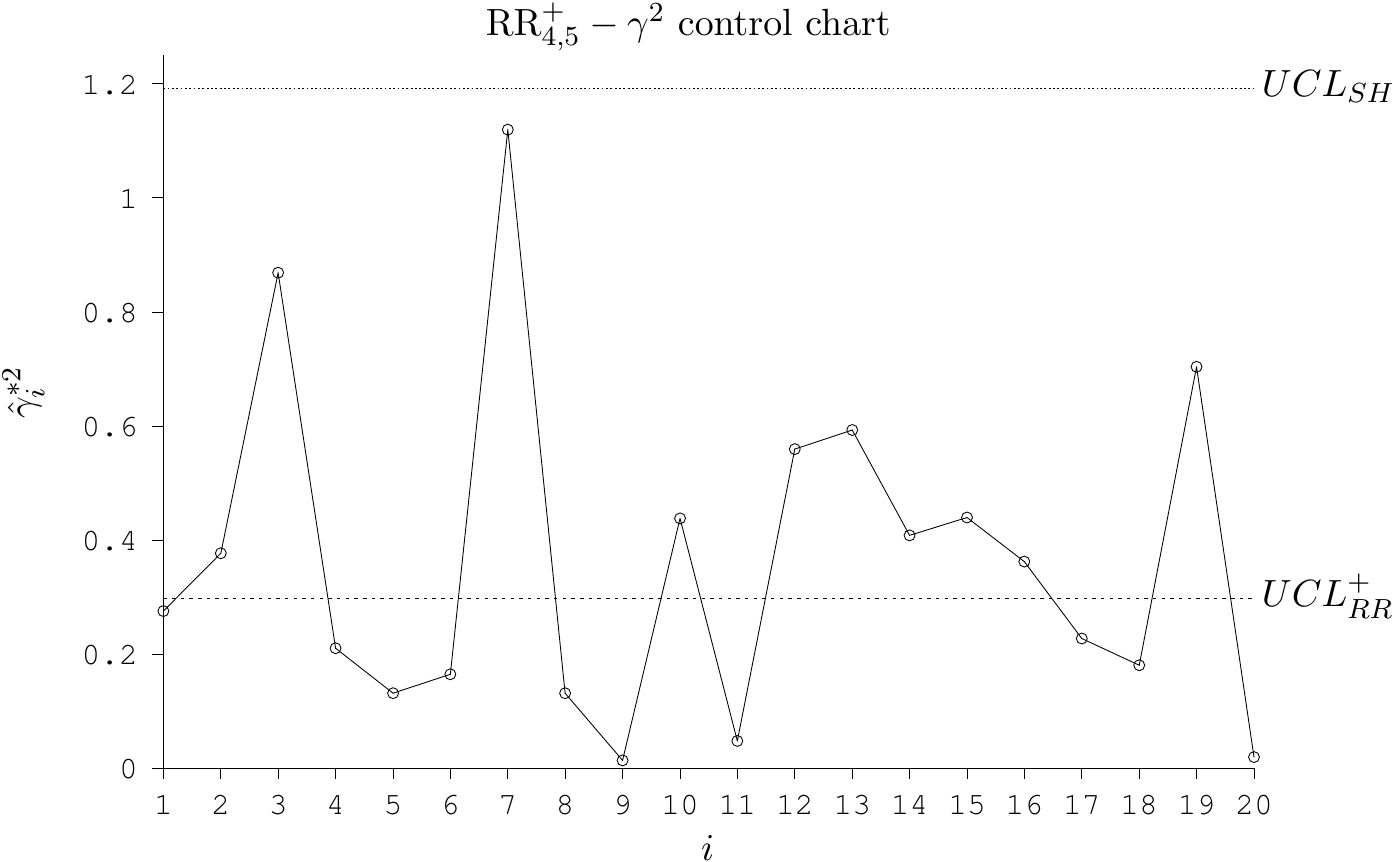} \\[5mm]
  \end{center}
  \caption{The upward CUSUM-$\gamma^2$ control chart in the presence of the measurement error corresponding to the Phase II data in Table \ref{tab:data}.}
\label{fig:vidu}
\end{figure}

According to \cite{Tran_2018_CVME} under the presence of the measurement error, we suppose that  the parameters of the linear covariate error model are $\eta=0.28$, $\theta=0.05$, $B=1$, and $m=1$. 
Based on the process engineer's experience, a specific shift size $\tau=1.25$ was expected to detect from the process. Therefore, we apply the upper-sided RR$_{r,s-\gamma^2}$ control chart to monitor the CV squared.
The control limits of the RR$^+_{2,3}-\gamma^2$, RR$^+_{3,4}-\gamma^2$ and  RR$_{4,5}-\gamma^2$ chart are found to be $UCL^+=0.5567$, $UCL^+=0.3821$ and $UCL^+=0.2972$, respectively. The values of $\gamma^{*2}_i$ are then plotted in  these charts (Figure \ref{fig:vidu}) long with the control limit $UCL^+$. For purpose of comparison, we also draw the control limit ($UCL^+=1.1913 $)of the original Shewhart control chart with the same parameters.

As can be seen from the Figure \ref{fig:vidu}, the RR$^+_{2,3}-\gamma^2$, RR$^+_{3,4}-\gamma^2$ and  RR$_{4,5}-\gamma^2$ chart signal the
occurrence of the out-of-control condition by two-out-of-three, three-out-of-four, and four-out-of-five (respectively) successive plotting points above the corresponding control limits from the sample \#12. Meanwhile, the Shewhart chart fails to detect this out-of-control condition.
% The process is allowed to continue after the signal. Thus, the corrective actions are started by the repair crew who find and eliminate the assignable cause after sample  and restore the process back to the in-control condition.  
  
\begin{center}
PLEASE INSERT TABLE  \ref{tab:data} HERE\\
PLEASE INSERT FIGURE  \ref{fig:vidu} HERE
\end{center}

%%%%%%%%%%%%%%%%%%%%%%%%%%%%%%%%%%%%%%%%%%%%%%%%%%%%%%%%%%%%%%%%%%%%%%%%%%%%%%%%
\section{Concluding remarks}
\label{sec:conclu}
In this paper,  the performance of Run Rules control charts is improved slightly by monitoring the CV squared with the two one-sided charts rather than monitoring directly the CV with a two-sided chart as in a previous study in the literature. The effect of measurement errors on the performance of the RR$_{r,s}-{\gamma}^2$ control charts using a linear covariate error model is also investigated. We have pointed out the negative effect of measurement errors on the proposed charts: the increase of $\eta$ and $\theta$ leads to the increase of $EARL$. Moreover, the obtained results show that measuring repeatedly is not an efficient method for limiting these unexpected effects HANH HEO. \\
%%%%%%%%%%%%%%%%%%%%%%%%%%%%%%%%%%%%%%%%%%%%%%%%%%%%%%%%%%%%%%%%%%%%%%%%%%%%%%%%
%\section*{Acknowledgements}
%The authors would like to thank the anonymous referees for their valuable suggestions that helped to improve the quality of the final manuscript. Research activities of Phuong Hanh Tran have been funded by Vietnam International Education Development - Project 911.
%\bibliography{SPC_Reference}

\begin{thebibliography}{10}
\providecommand{\MR}{\relax\unskip\space MR }
\providecommand{\url}[1]{\normalfont{#1}}
\providecommand{\urlprefix}{Available at }

\bibitem{breunig2001}
R. Breunig, \emph{An almost unbiased estimator of the coefficient of
  variation}, Economics Letters 70 (2001), pp. 15--19.

\bibitem{Castagliola_Run-Rules_CV_2013}
P. Castagliola, A. Achoure, H. Taleb, G. Celano, and S. Psarakis,
  \emph{Monitoring the coefficient of variation using control charts with run
  rules}, Quality Technology \& Quantitative Management 10 (2013), pp. 75--94.

\bibitem{CastagliolaVSI_CV_2013}
P. Castagliola, A. Achouri, H. Taleb, G. Celano, and S. Psarakis,
  \emph{Monitoring the coefficient of variation using a variable sampling
  interval control chart}, Quality and Reliability Engineering International 29
  (2013), pp. 1135--1149.

\bibitem{Castagliola2015_VSS_CV}
P. Castagliola, A. Achouri, H. Taleb, G. Celano, and S. Psarakis,
  \emph{Monitoring the coefficient of variation using a variable sample size
  control chart}, The International Journal of Advanced Manufacturing
  Technology 81 (2015), pp. 1561--1576.

\bibitem{Castagliola_EWMA_CV_2011}
P. Castagliola, A. Amdouni, H. Taleb, and G. Celano, \emph{Monitoring the
  coefficient of variation using ewma charts}, Journal of Quality Technology 43
  (2011), pp. 249--265.

\bibitem{Khaw_VSSI_CV_2017}
K.W. Khaw, M.B.C. Khoo, W.C. Yeong, and Z. Wu, \emph{Monitoring the coefficient
  of variation using a variable sample size and sampling interval control
  chart}, Communications in Statistics - Simulation and Computation 46 (2017),
  pp. 5722--5794.

\bibitem{Muhammada_2018_VSSEWMA_CV}
A.N.B. Muhammada, W. Yeong, Z. Chonga, S. Limc, and M.B.C. Khood,
  \emph{Monitoring the coefcient of variation using a variable sample size ewma
  chart}, Computers \& Industrial Engineering 126 (2018), pp. 378--398.

\bibitem{Nguyen_2019_VSICV_ME}
H.D. Nguyen, Q.T. Nguyen, K.P. Tran, and D.P. Ho, \emph{On the performance of
  {VSI} shewhart control chart for monitoring the coefficient of variation in
  the presence of measurement errors}, The International Journal of Advanced
  Manufacturing Technology  (2019), pp. 1--33.
\bibitem{linna2001effect}
K.W. Linna and W.H. Woodall, \emph{Effect of measurement error on Shewhart control charts}, Journal of Quality technology 33 (2001),
  pp. 213--222.
\bibitem{Tran2017Runrules_median_mean}
K.P. Tran, \emph{Run rules median control charts for monitoring process mean in
  manufacturing}, Quality and Reliability Engineering International  (2017).

\bibitem{Tran2016_Runrules_RZ}
K.P. Tran, P. Castagliola, and G. Celano, \emph{Monitoring the {R}atio of {T}wo
  {N}ormal {V}ariables {U}sing {R}un {R}ules {T}ype {C}ontrol {C}harts},
  International Journal of Production Research 54 (2016), pp. 1670--1688.

\bibitem{Tran_2018_CVME}
K.P. Tran, C. Heuchenne, N. Balakrishnan, and M.B.C. Khoo, \emph{On the
  performance of coefficient of variation charts in the presence of measurement
  errors}, Quality and Reliability Engineering International In press (2018).

\bibitem{Tran_2018_runrule_t}
K. Tran, \emph{Designing of run rules t control charts for monitoring changes
  in the process mean}, Chemometrics and Intelligent Laboratory Systems Inpress
  (2018).
\bibitem{tran2019performance}
K.P. Tran, H.D. Nguyen, P.H. Tran and C. Heuchenne, \emph{On the performance of CUSUM control charts for monitoring the coefficient of variation with measurement errors}, The International Journal of Advanced Manufacturing Technology (2019), pp. 1--15
\bibitem{tran2019synthetic}
 P.H. Tran, K.P. Tran and R. Athanasios
 , \emph{A Synthetic median control chart for monitoring the process mean with measurement errors}, Quality and Reliability Engineering International (2019), pp. 1100-1116
\bibitem{ye2018novel}
J. Ye, P. Feng, C. Xu, Y. Ma, and S. Huang, \emph{A novel approach for chatter
  online monitoring using coefficient of variation in machining process}, The
  International Journal of Advanced Manufacturing Technology (2018), pp.
  1--11.

\bibitem{Yeong_VSI_CV_Direct_2017}
W.C. Yeong, M.B.C. Khoo, S.L. Lim, and M.H. Lee, \emph{A direct procdedure for
  monitoring the coefficient of variation using a variable sample size sheme},
  Communications in Statistics - Simulation and Computation 46 (2017), pp.
  4210--4225.

\bibitem{Yeong_CV_ME_2017}
W.C. Yeong, M.B.C. Khoo, S.L. Lim, and W.L. Teoh, \emph{The coefficient of
  variation chart with measurement error}, Quality Technology \& Quantitative
  Management  (2017), pp. 1--25.

\end{thebibliography}

\newpage
%%%%Fig1%%%%
%%%%Fig2%%%
%%%%Fig3%%%
%%%%Fig4%%%
%%%%Fig5%%%
%%%%Fig6%%%
%%%%Fig7%%%

%%%%Fig8%%%

%%%%%%%
\newpage
%%%%%%%

%%%Tabl1%%%%
\begin{table}

  \scalebox{1}{ \begin{tabular}{ccc|cc|cc|cc}
      \hline
      $\eta$ & $\theta$ & $\gamma_0$ &  $n=10$& $n=15$&$n=10$& $n=15$&$n=10$& $n=15$ \\
      \hline
      &&& \multicolumn{2}{c|}{RR$_{2,3}-\gamma^2$}& \multicolumn{2}{c|}{RR$_{3,4}-\gamma^2$}& \multicolumn{2}{c}{RR$_{3,4}-\gamma^2$} \\

              \hline 
 0.1& $0.01$ & $0.05$& $0.0004$&  $0.0011$ & $0.0007$  & $0.0014$& $0.0009$  & $0.0016$ \\
       &        &       & $0.0063$&  $0.0044$  & $0.0047$  & $0.0037$ & $0.0039$  & $0.0033$ \\
 &  & $0.10$ &$0.0015$&  $0.0043$ & $0.0027$   & $0.0055$ & $0.0038$  & $0.0065$ \\
       &        &        &$0.0254$&  $0.0175$ & $0.0191$  & $0.0148$  & $0.0156$  & $0.0132$\\
% &  & $0.15$ &$0.0034$& $0.0097$ & $0.0060$  & $0.0124$ & $0.0085$ & $0.0145$ \\
       &        &        &$0.0582$& $0.0399$ & $0.0435$  & $0.0336$ & $0.0354$  & $0.0299$\\
 &  & $0.20$ &$0.0059$&  $0.0172$ & $0.0107$  & $0.0220$ & $0.0151$  & $0.0257$ \\
       &        &        &$0.1061$& $0.0718$  & $0.0786$  & $0.0602$ & $0.0636$  & $0.0534$ \\
       \hline
$0.28$ & $0.05$ & $0.05$ &$0.0004$& $0.0011$ & $0.0007$ & $0.0014$  & $0.0009$  & $0.0016$ \\
       &        &        &$0.0062$& $0.0043$ & $0.0047$  & $0.0036$  & $0.0038$  & $0.0033$\\
 &  & $0.10$ &$0.0015$& $0.0043$ & $0.0027$  & $0.0055$ & $0.0037$  & $0.0064$ \\
       &        &        &$0.0251$& $0.0173$ & $0.0189$  & $0.0146$ & $0.0154$  & $0.0130$ \\
% & & $0.15$ &$0.0033$& $0.0096$ & $0.0060$  & $0.0123$ & $0.0084$  & $0.0143$ \\
       &        &        &$0.0575$& $0.0394$ & $0.0430$ & $0.0332$ & $0.0350$  & $0.0295$ \\
&  & $0.20$ &$0.0059$& $0.0170$ & $0.0106$  & $0.0218$ & $0.0149$ & $0.0254$ \\
       &        &        &$0.1047$& $0.0709$ & $0.0776$  & $0.0595$ & $0.0629$  & $0.0528$\\
       \hline
  \end{tabular}}
    \caption{Values of $LCL$ (first row) and $UCL$ (second row) for the RR$_{r,s}-\gamma^2$ control charts in the presence of measurement errors, for different values of $\eta$, $\theta$, $n$, $\gamma_0$, $B=1$ and $m=1$.}
\label{tab:LUCL}
\end{table}
%%%Tabl2%%%%%%%%%%%%%%%%%%%%%%
\begin{table}
\hspace{-2.5mm}
  \scalebox{0.8}{
    \begin{tabular}{|c|c|c|c|c|c|c|}
\hline
\multirow{2}{*}{ Charts}&  \multicolumn{2}{c|}{$\gamma_0=0.05$}&\multicolumn{2}{c|}{$\gamma_0=0.1$}&\multicolumn{2}{c|}{$\gamma_0=0.2$}\\
\cline{2-7}
&$n=5$& $n=15$ &$n=5$& $n=15$ & $n=5$& $n=15$\\
\hline
 RR$_{2,3}-{\gamma}^2$  & $(1.194,2.167)$ & $(1.487,2.010)$ &  $(1.170,2.183)$  & $(1.464,2.023)$ & $(1.088,2.245)$  & $(1.407,2.069)$ \\
 RR$_{3,4}-{\gamma}^2$& $(1.023,1.293)$ &  $(1.159,1.298)$ & $(1.003,1.301)$  & $(1.143,1.306)$ & $(0.930,1.331)$  & $(1.098,1.333)$ \\
 RR$_{4,5}-{\gamma}^2$& $(0.866,0.801)$  & $(0.915,0.872)$ & $(0.849,0.808)$  & $(0.902,0.878)$ & $(0.785,0.832)$  & $(0.865,0.899)$\\

 \hline
    \end{tabular}}
 \caption{Values of the parameters $k_d$ (left side) and $k_u$ (right size) of the downward chart and the upward RR$_{r,s}-{\gamma}^2$ charts for $\gamma_0 = \{0.05, 0.1, 0.2\}$ and $n=\{5,15\}$ when $ARL_0=370.4$.}
  \label{tab:nome}
  \end{table}
%%%Tabl3%%%%
\begin{table}
\begin{center}
\hspace{-3mm}
  \scalebox{0.8}{
       \begin{tabular}{|c|c|c|c|c|c|c|c|}
\hline
\multirow{2}{*}{Charts}&\multirow{2}{*}{$\tau$}&  \multicolumn{2}{c|}{$\gamma_0=0.05$}&  \multicolumn{2}{c|}{$\gamma_0=0.1$}&\multicolumn{2}{c|}{$\gamma_0=0.2$}\\
\cline{3-8}
&&$n=5$& $n=15$ & $n=5$& $n=15$& $n=5$& $n=15$\\
\hline
 RR$_{2,3}^--{\gamma}^2$&\multirow{3}{*}{0.5}& $(8.1,6.6)$ &  $(2.1,0.3)$  & $(8.1,6.5)$  & $(2.1,0.3)$& $(8.2,6.7)$  & $(2.1,0.3)$ \\
 RR$_{3,4}^--{\gamma}^2$&& $(5.6,3.3)$  & $(3.0,0.1)$ & $(5.6,3.3)$  & $(3.0,0.1)$& $(5.7,3.4)$  & $(3.0,0.1)$ \\
 RR$_{4,5}^--{\gamma}^2$&& $(5.4,2.1)$ &  $(4.0,0.1)$ & $(5.4,2.1)$ & $(4.0,0.1)$& $(5.4,2.2)$  & $(4.0,0.1)$ \\
 \hline
 RR$_{2,3}^--{\gamma}^2$&\multirow{3}{*}{0.65}& $(26.9,25.2)$  & $(3.8,2.3)$& $(26.6,25.0)$  & $(3.8,2.2)$ & $(27.1,25.5)$  & $(3.9,2.4)$ \\
 RR$_{3,4}^--{\gamma}^2$&& $((16.2,13.7)$  & $(3.8,1.4)$ & $(16.2,13.8)$ & $(3.8,1.4)$ & $(16.6,14.1)$  & $(3.9,1.5)$ \\
  RR$_{4,5}^--{\gamma}^2$&& $(12.6,9.5)$  & $(4.5,1.0)$ & $(12.7,9.6)$  & $(4.5,1.0)$& $(13.0,9.9)$  & $(4.5,1.0)$ \\
 \hline
 RR$_{2,3}^--{\gamma}^2$ &\multirow{3}{*}{0.8}& $(87.9,86.1)$  & $(17.9,16.2)$ & $(87.2,85.4)$  & $(17.5,15.9)$ & $(88.4,86.6)$  & $(18.2,16.5)$\\
 RR$_{3,4}^--{\gamma}^2$&& $(59.8,57.1)$  & $(12.6,10.2)$ & $(59.9,57.2)$  & $(12.7,10.3)$& $(61.1,58.4)$  & $(13.2,10.8)$ \\
 RR$_{4,5}^--{\gamma}^2$&& $(46.8,43.4)$ &  $(11.1,8.0)$ & $(47.0,43.6)$  & $(11.2,8.0)$  & $(48.1,44.7)$  & $(11.6,8.4)$ \\
 \hline
 RR$_{2,3}^--{\gamma}^2$ &\multirow{3}{*}{0.9}& $(184.4,182.5)$  & $(75.4,73.7)$ & $(183.7,181.9)$ & $(73.9,72.2)$ & $(185.2,183.4)$  & $(75.7,73.9)$ \\
 RR$_{3,4}^--{\gamma}^2$&& $(149.3,146.5)$  & $(55.4,52.7)$ & $(149.5,146.7)$ &  $(55.3,52.6)$ & $(151.3,148.5)$ &  $(57.1,54.5)$ \\
 RR$_{4,5}^--{\gamma}^2$&& $(128.7,125.1)$  & $(46.5,43.1)$ & $(129.1,125.4)$ &  $(46.7,43.3)$& $(130.9,127.3)$ &  $(48.3,44.9)$ \\
 \hline
 RR$_{2,3}^+-{\gamma}^2$ &\multirow{3}{*}{1.10}& $(95.9,94.1)$ &  $(45.8,44.1)$ & $(96.5,94.7)$  & $(46.4,44.6)$ & $(98.7,96.9)$  & $(48.5,46.8)$ \\
 RR$_{3,4}^+-{\gamma}^2$&& $(94.2,91.5)$  & $(42.3,39.7)$ & $(94.8,92.0)$ &  $(42.8,40.1)$ & $(97.0,94.2)$  & $(44.7,42.1)$ \\
 RR$_{4,5}^+-{\gamma}^2$&& $(94.9,91.4)$  & $(41.3,37.9)$ & $(95.4,91.9)$ &  $(41.7,38.3)$ & $(97.6,94.0)$  & $(43.6,40.2)$ \\
 \hline
RR$_{2,3}^+-{\gamma}^2$ &\multirow{3}{*}{1.25}& $(25.8,24.2)$  & $(8.7,7.1)$ & $(26.1,24.4)$  & $(8.8,7.2)$ & $(27.2,25.6)$  & $(9.4,7.8)$ \\
 RR$_{3,4}^+-{\gamma}^2$&& $(26.3,23.8)$  & $(8.9,6.5)$ & $(26.5,24.0)$ &  $(9.0,6.6)$ & $(27.6,25.1)$  & $(9.5,7.1)$ \\
 RR$_{4,5}^+-{\gamma}^2$&& $(27.5,24.2)$  & $(9.5,6.4)$ & $(27.8,24.5)$  & $(9.6,6.5)$ & $(28.9,25.5)$  & $(10.1,7.0)$ \\
 \hline
RR$_{2,3}^+-{\gamma}^2$ &\multirow{3}{*}{1.5}& $(8.1,6.6)$  & $(3.1,1.5)$ & $(8.2,6.7)$  & $(3.1,1.6)$ & $(8.6,7.1)$ & $(3.3,1.7)$ \\
 RR$_{3,4}^+-{\gamma}^2$&& $(9.1,6.7)$  & $(3.9,1.4)$ & $(9.2,6.8)$ &  $(3.9,1.5)$ & $(9.6,7.2)$ &  $(4.0,1.6)$ \\
 RR$_{4,5}^+-{\gamma}^2$&& $(10.2,7.1)$  & $(4.8,1.4)$ & $(10.3,7.2)$ &  $(4.8,1.4)$ & $(10.7,7.6)$ &  $(4.9,1.6)$ \\
 \hline
RR$_{2,3}^+-{\gamma}^2$ &\multirow{3}{*}{2.0}& $(3.4,1.9)$  & $(2.1,0.3)$ &  $(3.4,1.9)$  & $(2.1,0.3)$& $(3.6,2.1)$  & $(2.1,0.4)$ \\
 RR$_{3,4}^+-{\gamma}^2$&& $(4.3,2.0)$  & $(3.1,0.3)$ & $(4.4,2.0)$ &  $(3.1,0.3)$ & $(4.6,2.2)$  & $(3.1,0.3)$ \\
 RR$_{4,5}^+-{\gamma}^2$&& $(5.3,2.1)$  & $(4.0,0.2)$ & $(5.4,2.1)$  & $(4.0,0.2)$ & $(5.6,2.3)$ &  $(4.1,0.3)$\\
\hline
    \end{tabular}}
    \end{center}
 \caption{Values of $(ARL_1,SDRL_1)$ of RR$_{r,s}-{\gamma}^2$ charts corresponding to the chart parameters in Table 1 for various situations of $\tau$.}
  \label{tab:ARL1}
  \end{table}

%%%Tabl4%%%%%%%%%%%%%%%%%%%%
\begin{table}
  %\hspace*{-20mm}
  \scalebox{0.65}{
    \begin{tabular}{cccccccc}
      Charts&$\tau$ & $\eta=0$ & $\eta=0.1$ & $\eta=0.2$ & $\eta=0.3$ & $\eta=0.5$& $\eta=1$ \\
\hline
          & \multicolumn{5}{c}{$n=5$} \\
\hline
\hline
 & \multicolumn{6}{c}{$n=5$} \\
\multirow{6}{*}{RR$_{2,3}-\gamma^2$}& $0.5$ &  $(8.92,8.85,8.98)$ &  $(8.94,8.85,8.99)$ &  $(8.92,8.84,8.99)$ &  $(8.92,8.86,9.01)$ &  $(8.90,8.84,9.06)$ &  $(8.87,8.87,9.24)$ \\
& $0.7$ &  $(29.38,29.11,29.54)$ &  $(29.45,29.14,29.58)$ &  $(29.35,29.06,29.57)$ &  $(29.40,29.13,29.65)$ &  $(29.29,29.09,29.80)$ &  $(29.19,29.18,30.42)$ \\
& $0.8$ &  $(93.12,92.52,93.41)$ &  $(93.28,92.44,93.58)$ &  $(93.15,92.24,93.38)$ &  $(93.20,92.54,93.68)$ &  $(92.90,92.39,93.98)$ &  $(92.74,92.60,95.22)$ \\
&$1.3$ &  $(28.57,28.83,29.93)$ &  $(28.57,28.84,29.94)$ &  $(28.57,28.85,29.99)$ &  $(28.58,28.87,30.06)$ &  $(28.59,28.92,30.30)$ &  $(28.66,29.19,31.45)$ \\
& $1.5$ &  $(9.07,9.17,9.62)$ &  $(9.07,9.17,9.62)$ &  $(9.07,9.18,9.64)$ &  $(9.07,9.19,9.67)$ &  $(9.07,9.21,9.77)$ &  $(9.10,9.32,10.24)$ \\
& $2.0$ &  $(3.70,3.74,3.92)$ &  $(3.70,3.74,3.92)$ &  $(3.70,3.74,3.92)$ &  $(3.70,3.75,3.94)$ &  $(3.70,3.76,3.98)$ &  $(3.71,3.80,4.16)$ \\
\hline
\multirow{6}{*}{RR$_{3,4}-\gamma^2$} & $0.5$ &  $(6.01,6.02,6.11)$ &  $(6.02,6.02,6.12)$ &  $(6.02,6.02,6.12)$ &  $(6.01,6.02,6.13)$ &  $(6.01,6.02,6.15)$ &  $(6.01,6.05,6.26)$ \\
& $0.7$ &  $(17.69,17.71,18.09)$ &  $(17.71,17.71,18.11)$ &  $(17.71,17.71,18.11)$ &  $(17.70,17.73,18.15)$ &  $(17.71,17.73,18.25)$ &  $(17.68,17.82,18.68)$ \\
 &$0.8$ &  $(64.10,64.21,65.33)$ &  $(64.20,64.19,65.35)$ &  $(64.22,64.17,65.37)$ &  $(64.13,64.24,65.47)$ &  $(64.22,64.25,65.78)$ &  $(64.07,64.48,67.06)$ \\
& $1.3$ &  $(28.92,29.17,30.21)$ &  $(28.92,29.17,30.22)$ &  $(28.92,29.18,30.26)$ &  $(28.93,29.20,30.34)$ &  $(28.93,29.25,30.56)$ &  $(29.00,29.51,31.63)$ \\
 &$1.5$ &  $(10.01,10.12,10.55)$ &  $(10.01,10.12,10.55)$ &  $(10.01,10.12,10.57)$ &  $(10.02,10.13,10.60)$ &  $(10.02,10.15,10.69)$ &  $(10.05,10.26,11.14)$ \\
 &$2.0$ &  $(4.67,4.71,4.89)$ &  $(4.67,4.71,4.89)$ &  $(4.67,4.72,4.90)$ &  $(4.67,4.72,4.91)$ &  $(4.68,4.73,4.95)$ &  $(4.69,4.77,5.13)$ \\
\hline
\multirow{6}{*}{RR$_{4,5}-\gamma^2$}& $0.5$ & $(5.64,5.65,5.71)$ & $(5.64,5.65,5.71)$ & $(5.64,5.65,5.72)$ & $(5.64,5.65,5.72)$ & $(5.64,5.65,5.74)$ & $(5.64,5.67,5.81)$ \\
& $0.7$ & $(13.75,13.79,14.09)$ & $(13.76,13.80,14.10)$ & $(13.76,13.81,14.12)$ & $(13.75,13.81,14.13)$ & $(13.76,13.81,14.20)$ & $(13.76,13.89,14.52)$ \\
& $0.8$ & $(50.44,50.60,51.63)$ & $(50.49,50.63,51.67)$ & $(50.51,50.66,51.74)$ & $(50.45,50.66,51.78)$ & $(50.46,50.67,52.03)$ & $(50.48,50.96,53.14)$ \\
& $1.3$ & $(30.17,30.42,31.45)$ & $(30.18,30.43,31.46)$ & $(30.19,30.43,31.51)$ & $(30.18,30.45,31.57)$ & $(30.20,30.51,31.80)$ & $(30.25,30.76,32.85)$ \\
& $1.5$ & $(11.17,11.27,11.70)$ & $(11.17,11.27,11.71)$ & $(11.17,11.27,11.73)$ & $(11.17,11.28,11.76)$ & $(11.17,11.30,11.85)$ & $(11.20,11.41,12.30)$ \\
& $2.0$ & $(5.68,5.72,5.90)$ & $(5.68,5.72,5.90)$ & $(5.68,5.72,5.91)$ & $(5.68,5.72,5.92)$ & $(5.68,5.73,5.96)$ & $(5.69,5.78,6.14)$ \\
\hline
\hline
 & \multicolumn{6}{c}{$n=15$} \\
\multirow{6}{*}{RR$_{2,3}-\gamma^2$} & $0.5$ &  $(2.12,2.12,2.13)$ &  $(2.12,2.12,2.13)$ &  $(2.12,2.12,2.13)$ &  $(2.12,2.12,2.13)$ &  $(2.12,2.12,2.14)$ &  $(2.12,2.12,2.15)$ \\
& $0.7$ &  $(4.12,4.08,4.18)$ &  $(4.12,4.08,4.18)$ &  $(4.12,4.08,4.18)$ &  $(4.11,4.08,4.19)$ &  $(4.11,4.08,4.22)$ &  $(4.09,4.10,4.35)$ \\
& $0.8$ &  $(19.86,19.42,19.97)$ &  $(19.83,19.43,20.02)$ &  $(19.83,19.45,20.02)$ &  $(19.80,19.45,20.06)$ &  $(19.73,19.43,20.20)$ &  $(19.56,19.55,21.01)$ \\
& $1.3$ &  $(9.68,9.81,10.41)$ &  $(9.68,9.82,10.41)$ &  $(9.68,9.82,10.44)$ &  $(9.68,9.83,10.48)$ &  $(9.69,9.86,10.61)$ &  $(9.72,10.01,11.22)$ \\
& $1.5$ &  $(3.32,3.36,3.52)$ &  $(3.32,3.36,3.52)$ &  $(3.32,3.36,3.53)$ &  $(3.32,3.36,3.54)$ &  $(3.32,3.37,3.58)$ &  $(3.33,3.41,3.75)$ \\
& $2.0$ &  $(2.13,2.14,2.17)$ &  $(2.13,2.14,2.17)$ &  $(2.13,2.14,2.18)$ &  $(2.13,2.14,2.18)$ &  $(2.13,2.14,2.19)$ &  $(2.13,2.15,2.23)$ \\
\hline
\multirow{6}{*}{RR$_{3,4}-\gamma^2$} & $0.5$ &  $(3.02,3.02,3.03)$ &  $(3.02,3.02,3.03)$ &  $(3.02,3.02,3.03)$ &  $(3.02,3.02,3.03)$ &  $(3.02,3.02,3.03)$ &  $(3.02,3.03,3.03)$ \\
& $0.7$ &  $(4.02,4.02,4.09)$ &  $(4.02,4.02,4.09)$ &  $(4.01,4.02,4.10)$ &  $(4.01,4.02,4.10)$ &  $(4.01,4.03,4.12)$ &  $(4.01,4.04,4.20)$ \\
& $0.8$ &  $(13.91,13.92,14.41)$ &  $(13.92,13.91,14.42)$ &  $(13.90,13.91,14.44)$ &  $(13.90,13.92,14.48)$ &  $(13.89,13.95,14.59)$ &  $(13.87,14.06,15.13)$ \\
& $1.3$ &  $(9.77,9.89,10.42)$ &  $(9.77,9.90,10.42)$ &  $(9.78,9.90,10.44)$ &  $(9.78,9.91,10.48)$ &  $(9.78,9.94,10.59)$ &  $(9.81,10.07,11.13)$ \\
& $1.5$ &  $(4.10,4.13,4.28)$ &  $(4.10,4.13,4.28)$ &  $(4.10,4.14,4.29)$ &  $(4.10,4.14,4.30)$ &  $(4.10,4.14,4.33)$ &  $(4.11,4.18,4.48)$ \\
 &$2.0$ &  $(3.09,3.10,3.12)$ &  $(3.09,3.10,3.13)$ &  $(3.09,3.10,3.13)$ &  $(3.09,3.10,3.13)$ &  $(3.09,3.10,3.14)$ &  $(3.09,3.11,3.17)$ \\
 \hline
\multirow{6}{*}{RR$_{4,5}-\gamma^2$} & $0.5$ & $(4.01,4.01,4.01)$ & $(4.01,4.01,4.01)$ & $(4.01,4.01,4.01)$ & $(4.01,4.01,4.01)$ & $(4.01,4.01,4.01)$ & $(4.01,4.01,4.01)$ \\
 &$0.7$ & $(4.58,4.59,4.64)$ & $(4.58,4.59,4.64)$ & $(4.58,4.59,4.64)$ & $(4.58,4.59,4.65)$ & $(4.58,4.59,4.66)$ & $(4.58,4.60,4.71)$ \\
& $0.8$ & $(12.09,12.14,12.55)$ & $(12.09,12.14,12.56)$ & $(12.09,12.14,12.58)$ & $(12.09,12.15,12.61)$ & $(12.09,12.17,12.70)$ & $(12.09,12.27,13.15)$ \\
 &$1.3$ & $(10.37,10.48,10.97)$ & $(10.37,10.48,10.98)$ & $(10.37,10.49,11.00)$ & $(10.37,10.49,11.03)$ & $(10.38,10.52,11.14)$ & $(10.40,10.64,11.64)$ \\
 &$1.5$ & $(4.97,5.00,5.13)$ & $(4.97,5.00,5.13)$ & $(4.97,5.00,5.14)$ & $(4.97,5.00,5.15)$ & $(4.97,5.01,5.18)$ & $(4.98,5.04,5.32)$ \\
& $2.0$ & $(4.07,4.08,4.10)$ & $(4.07,4.08,4.10)$ & $(4.07,4.08,4.10)$ & $(4.07,4.08,4.10)$ & $(4.07,4.08,4.11)$ & $(4.07,4.08,4.13)$ \\              
      \hline
    \end{tabular}}
    \caption{The $ARL$ values of the RR$_{r,s}-\gamma^2$ control charts in the presence of measurement errors 
     for $\gamma_0=0.05$ (left side), $\gamma_0=0.1$ (middle) and $\gamma_0=0.2$ (right side),  and
     for different values of $\eta$, $\theta=0.05$, $\tau$, $n$, $B=1$, $m=1$.}  
  \label{tab:eta}
\end{table}

%%%Tabl5%%%%

\begin{table}
% \vspace*{20mm}
  \scalebox{0.65}{
    \begin{tabular}{cccccccc}
          Charts&$\tau$ & $\theta=0$ & $\theta=0.01$ & $\theta=0.02$ & $\theta=0.03$ & $\theta=0.04$& $\theta=0.05$ \\
\hline
          & \multicolumn{5}{c}{$n=5$} \\
\hline
\hline
& \multicolumn{6}{c}{$n=5$} \\
\multirow{6}{*}{RR$_{2,3}-\gamma^2$} & $0.5$ & $(8.11,8.06,8.22)$ & $(8.28,8.21,8.38)$ & $(8.43,8.35,8.53)$ & $(8.59,8.52,8.69)$ & $(8.75,8.69,8.84)$ & $(8.93,8.85,9.01)$ \\
&$0.7$ & $(26.88,26.65,27.19)$ & $(27.38,27.15,27.71)$ & $(27.87,27.55,28.15)$ & $(28.36,28.10,28.66)$ & $(28.87,28.62,29.12)$ & $(29.42,29.13,29.64)$ \\
&$0.8$ & $(87.72,87.23,88.36)$ & $(88.96,88.30,89.56)$ & $(89.94,89.12,90.51)$ & $(90.85,90.33,91.69)$ & $(91.95,91.30,92.44)$ & $(93.24,92.53,93.63)$ \\
&$1.3$ & $(25.84,26.13,27.35)$ & $(26.38,26.67,27.88)$ & $(26.93,27.21,28.42)$ & $(27.47,27.76,28.96)$ & $(28.02,28.31,29.50)$ & $(28.58,28.86,30.04)$ \\
&$1.5$ & $(8.09,8.20,8.68)$ & $(8.28,8.39,8.87)$ & $(8.47,8.59,9.07)$ & $(8.67,8.78,9.26)$ & $(8.87,8.98,9.46)$ & $(9.07,9.18,9.66)$ \\
&$2.0$ & $(3.38,3.42,3.61)$ & $(3.44,3.48,3.67)$ & $(3.50,3.55,3.73)$ & $(3.57,3.61,3.80)$ & $(3.63,3.68,3.87)$ & $(3.70,3.75,3.93)$ \\
\hline
\multirow{6}{*}{RR$_{3,4}-\gamma^2$} &$0.5$ & $(5.60,5.61,5.71)$ & $(5.68,5.69,5.80)$ & $(5.76,5.77,5.88)$ & $(5.84,5.85,5.96)$ & $(5.93,5.93,6.04)$ & $(6.01,6.02,6.12)$ \\
&$0.7$ & $(16.15,16.20,16.62)$ & $(16.47,16.50,16.93)$ & $(16.76,16.78,17.24)$ & $(17.07,17.10,17.53)$ & $(17.38,17.40,17.84)$ & $(17.70,17.72,18.14)$ \\
&$0.8$ & $(59.75,59.92,61.22)$ & $(60.70,60.79,62.12)$ & $(61.56,61.53,63.02)$ & $(62.38,62.49,63.79)$ & $(63.26,63.35,64.67)$ & $(64.20,64.24,65.45)$ \\
&$1.3$ & $(26.28,26.56,27.72)$ & $(26.80,27.08,28.23)$ & $(27.33,27.60,28.75)$ & $(27.86,28.13,29.27)$ & $(28.39,28.66,29.79)$ & $(28.92,29.19,30.31)$ \\
&$1.5$ & $(9.05,9.16,9.63)$ & $(9.24,9.35,9.82)$ & $(9.43,9.54,10.01)$ & $(9.62,9.73,10.20)$ & $(9.82,9.93,10.39)$ & $(10.02,10.13,10.59)$ \\
&$2.0$ & $(4.35,4.39,4.58)$ & $(4.41,4.46,4.64)$ & $(4.48,4.52,4.70)$ & $(4.54,4.58,4.77)$ & $(4.61,4.65,4.84)$ & $(4.67,4.72,4.91)$ \\
\hline
\multirow{6}{*}{RR$_{4,5}-\gamma^2$} &$0.5$ & $(5.37,5.39,5.46)$ & $(5.43,5.44,5.51)$ & $(5.48,5.49,5.56)$ & $(5.53,5.54,5.61)$ & $(5.58,5.59,5.67)$ & $(5.64,5.65,5.72)$ \\
&$0.7$ & $(12.65,12.70,13.04)$ & $(12.87,12.92,13.25)$ & $(13.09,13.14,13.47)$ & $(13.31,13.35,13.68)$ & $(13.53,13.58,13.90)$ & $(13.76,13.80,14.13)$ \\
&$0.8$ & $(46.80,47.00,48.23)$ & $(47.56,47.75,48.90)$ & $(48.26,48.47,49.65)$ & $(49.00,49.14,50.35)$ & $(49.70,49.91,51.07)$ & $(50.47,50.63,51.74)$ \\
&$1.3$ & $(27.55,27.82,28.97)$ & $(28.07,28.34,29.48)$ & $(28.60,28.86,29.99)$ & $(29.13,29.39,30.51)$ & $(29.65,29.92,31.04)$ & $(30.19,30.44,31.56)$ \\
&$1.5$ & $(10.18,10.30,10.77)$ & $(10.38,10.49,10.96)$ & $(10.57,10.68,11.15)$ & $(10.77,10.88,11.35)$ & $(10.97,11.08,11.55)$ & $(11.17,11.28,11.75)$ \\
&$2.0$ & $(5.34,5.39,5.58)$ & $(5.41,5.45,5.64)$ & $(5.47,5.52,5.71)$ & $(5.54,5.58,5.77)$ & $(5.61,5.65,5.84)$ & $(5.68,5.72,5.91)$ \\
\hline
\hline
& \multicolumn{6}{c}{$n=15$} \\
\multirow{6}{*}{RR$_{2,3}-\gamma^2$}  &$0.5$ & $(2.09,2.09,2.10)$ & $(2.09,2.09,2.10)$ & $(2.10,2.10,2.11)$ & $(2.11,2.11,2.12)$ & $(2.11,2.11,2.13)$ & $(2.12,2.12,2.13)$ \\
&$0.7$ & $(3.79,3.76,3.87)$ & $(3.85,3.82,3.94)$ & $(3.92,3.89,4.00)$ & $(3.98,3.95,4.06)$ & $(4.04,4.02,4.12)$ & $(4.12,4.08,4.19)$ \\
&$0.8$ & $(17.85,17.59,18.24)$ & $(18.22,17.94,18.60)$ & $(18.65,18.32,18.96)$ & $(19.00,18.68,19.31)$ & $(19.39,19.08,19.68)$ & $(19.82,19.40,20.05)$ \\
&$1.3$ & $(8.66,8.81,9.44)$ & $(8.86,9.01,9.64)$ & $(9.06,9.21,9.84)$ & $(9.26,9.41,10.05)$ & $(9.47,9.62,10.26)$ & $(9.68,9.83,10.47)$ \\
&$1.5$ & $(3.07,3.11,3.28)$ & $(3.12,3.16,3.33)$ & $(3.17,3.21,3.38)$ & $(3.22,3.26,3.43)$ & $(3.27,3.31,3.48)$ & $(3.32,3.36,3.54)$ \\
&$2.0$ & $(2.08,2.09,2.13)$ & $(2.09,2.10,2.14)$ & $(2.10,2.11,2.15)$ & $(2.11,2.12,2.16)$ & $(2.12,2.13,2.17)$ & $(2.13,2.14,2.18)$ \\
\hline
\multirow{6}{*}{RR$_{3,4}-\gamma^2$}  &$0.5$ & $(3.01,3.02,3.02)$ & $(3.02,3.02,3.02)$ & $(3.02,3.02,3.02)$ & $(3.02,3.02,3.02)$ & $(3.02,3.02,3.03)$ & $(3.02,3.02,3.03)$ \\
&$0.7$ & $(3.83,3.84,3.91)$ & $(3.87,3.87,3.95)$ & $(3.90,3.91,3.99)$ & $(3.94,3.95,4.02)$ & $(3.98,3.98,4.06)$ & $(4.02,4.02,4.10)$ \\
&$0.8$ & $(12.64,12.68,13.22)$ & $(12.89,12.92,13.47)$ & $(13.14,13.16,13.71)$ & $(13.38,13.41,13.97)$ & $(13.64,13.67,14.22)$ & $(13.91,13.92,14.46)$ \\
&$1.3$ & $(8.87,9.00,9.56)$ & $(9.04,9.18,9.74)$ & $(9.22,9.36,9.92)$ & $(9.41,9.54,10.10)$ & $(9.59,9.72,10.28)$ & $(9.78,9.91,10.47)$ \\
&$1.5$ & $(3.88,3.91,4.06)$ & $(3.92,3.96,4.11)$ & $(3.97,4.00,4.15)$ & $(4.01,4.04,4.20)$ & $(4.06,4.09,4.25)$ & $(4.10,4.14,4.29)$ \\
&$2.0$ & $(3.06,3.06,3.09)$ & $(3.07,3.07,3.10)$ & $(3.07,3.08,3.10)$ & $(3.08,3.08,3.11)$ & $(3.09,3.09,3.12)$ & $(3.09,3.10,3.13)$ \\
\hline
\multirow{6}{*}{RR$_{4,5}-\gamma^2$}  &$0.5$ & $(4.00,4.00,4.00)$ & $(4.00,4.00,4.01)$ & $(4.00,4.00,4.01)$ & $(4.01,4.01,4.01)$ & $(4.01,4.01,4.01)$ & $(4.01,4.01,4.01)$ \\
&$0.7$ & $(4.46,4.47,4.52)$ & $(4.48,4.49,4.54)$ & $(4.51,4.51,4.57)$ & $(4.53,4.54,4.59)$ & $(4.55,4.56,4.62)$ & $(4.58,4.59,4.64)$ \\
&$0.8$ & $(11.10,11.17,11.62)$ & $(11.29,11.36,11.82)$ & $(11.48,11.55,12.01)$ & $(11.69,11.75,12.21)$ & $(11.89,11.94,12.40)$ & $(12.09,12.14,12.60)$ \\
&$1.3$ & $(9.50,9.63,10.15)$ & $(9.67,9.79,10.32)$ & $(9.84,9.97,10.49)$ & $(10.02,10.14,10.67)$ & $(10.19,10.31,10.85)$ & $(10.37,10.49,11.02)$ \\
&$1.5$ & $(4.77,4.80,4.93)$ & $(4.81,4.84,4.97)$ & $(4.85,4.88,5.02)$ & $(4.89,4.92,5.06)$ & $(4.93,4.96,5.10)$ & $(4.97,5.00,5.15)$ \\
&$2.0$ & $(4.05,4.05,4.07)$ & $(4.05,4.05,4.07)$ & $(4.06,4.06,4.08)$ & $(4.06,4.06,4.09)$ & $(4.07,4.07,4.09)$ & $(4.07,4.08,4.10)$ \\
            
      \hline
    \end{tabular}}
    \caption{The $ARL$ values of the RR$_{r,s}-\gamma^2$ control charts in the 
   presence of measurement errors for $\gamma_0=0.05$ (left side), $\gamma_0=0.1$ (middle) and $\gamma_0=0.2$ (right side),  and
   for different values of $\theta$, $\eta=0.28$, $\tau$, $n$, $B=1$, $m=1$.}  
  \label{tab:theta}
\end{table}

%%%Tabl6%%%2
\begin{table}
%\hspace{20mm}
  \scalebox{0.65}{
    \begin{tabular}{ccccccc}
      & \multicolumn{5}{c}{$n=5$} \\
  \hline
      Charts&$\tau$ & $B=0.8$ & $B=0.9$ & $B=1.0$ & $B=1.1$ & $B=1.2$ \\
      \hline
      \hline
& \multicolumn{5}{c}{$n=5$} \\
\multirow{6}{*}{RR$_{2,3}-\gamma^2$}  &0.5&$(9.13,9.05,9.22)$ & $(9.03,8.93,9.11)$ & $(8.93,8.85,9.01)$ & $(8.84,8.76,8.94)$ & $(8.79,8.72,8.87)$ \\
&0.7&$(30.00,29.74,30.26)$ & $(29.73,29.33,29.95)$ & $(29.42,29.13,29.64)$ & $(29.10,28.82,29.42)$ & $(28.99,28.71,29.24)$ \\
&0.8&$(94.27,93.68,94.95)$ & $(93.89,92.84,94.31)$ & $(93.24,92.53,93.63)$ & $(92.45,91.72,93.28)$ & $(92.23,91.63,92.81)$ \\
&1.3&$(29.28,29.57,30.79)$ & $(28.89,29.17,30.37)$ & $(28.58,28.86,30.04)$ & $(28.33,28.60,29.77)$ & $(28.11,28.39,29.55)$ \\
&1.5&$(9.33,9.44,9.95)$ & $(9.18,9.30,9.79)$ & $(9.07,9.18,9.66)$ & $(8.98,9.09,9.56)$ & $(8.90,9.01,9.48)$ \\
&2.0&$(3.79,3.83,4.03)$ & $(3.74,3.78,3.98)$ & $(3.70,3.75,3.93)$ & $(3.67,3.71,3.90)$ & $(3.64,3.69,3.87)$ \\
\hline
\multirow{6}{*}{RR$_{3,4}-\gamma^2$}  &$0.5$ & $(6.12,6.13,6.24)$ & $(6.06,6.06,6.18)$ & $(6.01,6.02,6.12)$ & $(5.98,5.98,6.09)$ & $(5.94,5.95,6.05)$ \\
&$0.7$ & $(18.08,18.11,18.55)$ & $(17.89,17.88,18.33)$ & $(17.70,17.72,18.14)$ & $(17.57,17.58,18.00)$ & $(17.44,17.46,17.88)$ \\
&$0.8$ & $(65.22,65.28,66.56)$ & $(64.70,64.64,66.01)$ & $(64.20,64.24,65.45)$ & $(63.85,63.84,65.08)$ & $(63.43,63.53,64.76)$ \\
&$1.3$ & $(29.60,29.88,31.04)$ & $(29.22,29.50,30.63)$ & $(28.92,29.19,30.31)$ & $(28.68,28.94,30.06)$ & $(28.48,28.74,29.84)$ \\
&$1.5$ & $(10.27,10.38,10.87)$ & $(10.13,10.24,10.71)$ & $(10.02,10.13,10.59)$ & $(9.93,10.03,10.49)$ & $(9.85,9.96,10.41)$ \\
&$2.0$ & $(4.76,4.81,5.00)$ & $(4.71,4.76,4.95)$ & $(4.67,4.72,4.91)$ & $(4.64,4.69,4.87)$ & $(4.62,4.66,4.84)$ \\
\hline
\multirow{6}{*}{RR$_{4,5}-\gamma^2$} &$0.5$ & $(5.71,5.72,5.79)$ & $(5.67,5.68,5.75)$ & $(5.64,5.65,5.72)$ & $(5.61,5.62,5.69)$ & $(5.59,5.60,5.67)$ \\
&$0.7$ & $(14.04,14.08,14.43)$ & $(13.88,13.93,14.26)$ & $(13.76,13.80,14.13)$ & $(13.66,13.70,14.02)$ & $(13.57,13.61,13.93)$ \\
&$0.8$ & $(51.38,51.54,52.76)$ & $(50.91,51.07,52.19)$ & $(50.47,50.63,51.74)$ & $(50.18,50.32,51.43)$ & $(49.86,50.02,51.15)$ \\
&$1.3$ & $(30.85,31.13,32.28)$ & $(30.48,30.75,31.88)$ & $(30.19,30.44,31.56)$ & $(29.94,30.20,31.30)$ & $(29.74,30.00,31.09)$ \\
&$1.5$ & $(11.42,11.54,12.03)$ & $(11.28,11.40,11.87)$ & $(11.17,11.28,11.75)$ & $(11.08,11.19,11.65)$ & $(11.00,11.11,11.57)$ \\
&$2.0$ & $(5.77,5.81,6.01)$ & $(5.72,5.76,5.96)$ & $(5.68,5.72,5.91)$ & $(5.65,5.69,5.88)$ & $(5.62,5.66,5.85)$ \\
\hline
\hline
& \multicolumn{5}{c}{$n=15$} \\
\multirow{6}{*}{RR$_{2,3}-\gamma^2$} &0.5&$(2.13,2.13,2.14)$ & $(2.13,2.13,2.14)$ & $(2.12,2.12,2.13)$ & $(2.12,2.12,2.13)$ & $(2.12,2.11,2.13)$ \\
&0.7&$(4.20,4.17,4.29)$ & $(4.15,4.12,4.23)$ & $(4.12,4.08,4.19)$ & $(4.08,4.05,4.16)$ & $(4.06,4.02,4.13)$ \\
&0.8&$(20.29,19.95,20.61)$ & $(20.05,19.69,20.26)$ & $(19.82,19.40,20.05)$ & $(19.64,19.25,19.86)$ & $(19.50,19.11,19.73)$ \\
&1.3&$(9.95,10.11,10.77)$ & $(9.80,9.95,10.60)$ & $(9.68,9.83,10.47)$ & $(9.58,9.73,10.36)$ & $(9.50,9.65,10.28)$ \\
&1.5&$(3.39,3.43,3.62)$ & $(3.35,3.39,3.57)$ & $(3.32,3.36,3.54)$ & $(3.30,3.34,3.51)$ & $(3.28,3.31,3.49)$ \\
&2.0&$(2.14,2.15,2.19)$ & $(2.13,2.14,2.18)$ & $(2.13,2.14,2.18)$ & $(2.12,2.13,2.17)$ & $(2.12,2.13,2.17)$ \\
\hline
\multirow{6}{*}{RR$_{3,4}-\gamma^2$}  &$0.5$ & $(3.03,3.03,3.03)$ & $(3.03,3.03,3.03)$ & $(3.02,3.02,3.03)$ & $(3.02,3.02,3.03)$ & $(3.02,3.02,3.03)$ \\
&$0.7$ & $(4.06,4.07,4.15)$ & $(4.04,4.04,4.12)$ & $(4.02,4.02,4.10)$ & $(4.00,4.00,4.08)$ & $(3.98,3.99,4.06)$ \\
&$0.8$ & $(14.22,14.25,14.82)$ & $(14.05,14.07,14.62)$ & $(13.91,13.92,14.46)$ & $(13.78,13.81,14.34)$ & $(13.68,13.71,14.23)$ \\
&$1.3$ & $(10.02,10.15,10.74)$ & $(9.88,10.02,10.59)$ & $(9.78,9.91,10.47)$ & $(9.69,9.82,10.38)$ & $(9.62,9.75,10.30)$ \\
&$1.5$ & $(4.16,4.20,4.36)$ & $(4.13,4.16,4.32)$ & $(4.10,4.14,4.29)$ & $(4.08,4.12,4.27)$ & $(4.06,4.10,4.25)$ \\
&$2.0$ & $(3.10,3.11,3.14)$ & $(3.10,3.10,3.13)$ & $(3.09,3.10,3.13)$ & $(3.09,3.09,3.12)$ & $(3.09,3.09,3.12)$ \\
\hline
\multirow{6}{*}{RR$_{4,5}-\gamma^2$}  &$0.5$ & $(4.01,4.01,4.01)$ & $(4.01,4.01,4.01)$ & $(4.01,4.01,4.01)$ & $(4.01,4.01,4.01)$ & $(4.01,4.01,4.01)$ \\
&$0.7$ & $(4.61,4.62,4.68)$ & $(4.59,4.60,4.66)$ & $(4.58,4.59,4.64)$ & $(4.57,4.58,4.63)$ & $(4.56,4.57,4.62)$ \\
&$0.8$ & $(12.34,12.40,12.88)$ & $(12.20,12.26,12.72)$ & $(12.09,12.14,12.60)$ & $(12.00,12.05,12.50)$ & $(11.92,11.98,12.42)$ \\
&$1.3$ & $(10.60,10.72,11.28)$ & $(10.47,10.60,11.14)$ & $(10.37,10.49,11.02)$ & $(10.29,10.41,10.93)$ & $(10.22,10.34,10.86)$ \\
&$1.5$ & $(5.03,5.06,5.21)$ & $(5.00,5.03,5.18)$ & $(4.97,5.00,5.15)$ & $(4.95,4.98,5.12)$ & $(4.94,4.97,5.11)$ \\
&$2.0$ & $(4.08,4.08,4.11)$ & $(4.08,4.08,4.11)$ & $(4.07,4.08,4.10)$ & $(4.07,4.07,4.10)$ & $(4.07,4.07,4.09)$ \\
 
      \hline
    \end{tabular}}
    \caption{The $ARL$ values of the RR$_{r,s}-\gamma^2$ control charts in the 
   presence of measurement errors for $\gamma_0=0.05$ (left side), $\gamma_0=0.1$ (middle) and $\gamma_0=0.2$ (right side),  and
   for different values of $B$, $\tau$, $n$,  $\eta=0.28$, $\theta=0.05$, $m=1$.}  
  \label{tab:B}
  \end{table}

%%%Tabl7%%%%
  \begin{table}
 % \vspace*{20mm}
  \scalebox{0.65}{
    \begin{tabular}{ccccccc}
      & \multicolumn{5}{c}{$n=5$} \\
   \hline
    Charts&  $\tau$ & $m=1$ & $m=3$ & $m=5$ & $m=7$ & $m=10$ \\
      \hline
      \hline
 & \multicolumn{5}{c}{$n=5$} \\
\multirow{6}{*}{RR$_{2,3}-\gamma^2$}  &$0.5$ & $(8.93,8.85,9.01)$ & $(8.92,8.84,8.99)$ & $(8.93,8.84,8.99)$ & $(8.94,8.85,8.99)$ & $(8.93,8.84,8.99)$ \\
&$0.7$ & $(29.42,29.13,29.64)$ & $(29.39,29.06,29.56)$ & $(29.41,29.09,29.56)$ & $(29.42,29.11,29.58)$ & $(29.39,29.07,29.58)$ \\
&$0.8$ & $(93.24,92.53,93.63)$ & $(93.00,92.20,93.47)$ & $(93.22,92.44,93.43)$ & $(93.20,92.50,93.57)$ & $(93.12,92.23,93.52)$ \\
&$1.3$ & $(28.58,28.86,30.04)$ & $(28.57,28.84,29.96)$ & $(28.57,28.84,29.95)$ & $(28.57,28.83,29.95)$ & $(28.57,28.83,29.94)$ \\
&$1.5$ & $(9.07,9.18,9.66)$ & $(9.07,9.18,9.63)$ & $(9.07,9.17,9.62)$ & $(9.07,9.17,9.62)$ & $(9.07,9.17,9.62)$ \\
&$2.0$ & $(3.70,3.75,3.93)$ & $(3.70,3.74,3.92)$ & $(3.70,3.74,3.92)$ & $(3.70,3.74,3.92)$ & $(3.70,3.74,3.92)$ \\
\hline
\multirow{6}{*}{RR$_{3,4}-\gamma^2$}  &$0.5$ & $(6.01,6.02,6.12)$ & $(6.02,6.02,6.12)$ & $(6.01,6.01,6.12)$ & $(6.02,6.02,6.12)$ & $(6.01,6.02,6.12)$ \\
&$0.7$ & $(17.70,17.72,18.14)$ & $(17.72,17.70,18.11)$ & $(17.70,17.69,18.11)$ & $(17.71,17.71,18.10)$ & $(17.69,17.71,18.11)$ \\
&$0.8$ & $(64.20,64.24,65.45)$ & $(64.26,64.15,65.33)$ & $(64.18,64.15,65.36)$ & $(64.18,64.21,65.35)$ & $(64.15,64.15,65.39)$ \\
&$1.3$ & $(28.92,29.19,30.31)$ & $(28.92,29.17,30.25)$ & $(28.92,29.17,30.23)$ & $(28.92,29.17,30.22)$ & $(28.92,29.17,30.22)$ \\
&$1.5$ & $(10.02,10.13,10.59)$ & $(10.01,10.12,10.56)$ & $(10.01,10.12,10.55)$ & $(10.01,10.12,10.55)$ & $(10.01,10.12,10.55)$ \\
&$2.0$ & $(4.67,4.72,4.91)$ & $(4.67,4.72,4.89)$ & $(4.67,4.71,4.89)$ & $(4.67,4.71,4.89)$ & $(4.67,4.71,4.89)$ \\
\hline
\multirow{6}{*}{RR$_{4,5}-\gamma^2$}  &$0.5$ & $(5.64,5.65,5.72)$ & $(5.64,5.65,5.71)$ & $(5.64,5.65,5.71)$ & $(5.64,5.65,5.71)$ & $(5.64,5.65,5.71)$ \\
&$0.7$ & $(13.76,13.80,14.13)$ & $(13.76,13.80,14.10)$ & $(13.75,13.80,14.10)$ & $(13.76,13.79,14.10)$ & $(13.75,13.79,14.10)$ \\
&$0.8$ & $(50.47,50.63,51.74)$ & $(50.50,50.65,51.67)$ & $(50.46,50.62,51.68)$ & $(50.48,50.58,51.67)$ & $(50.47,50.60,51.67)$ \\
&$1.3$ & $(30.19,30.44,31.56)$ & $(30.18,30.43,31.49)$ & $(30.18,30.43,31.47)$ & $(30.18,30.42,31.46)$ & $(30.17,30.42,31.46)$ \\
&$1.5$ & $(11.17,11.28,11.75)$ & $(11.17,11.27,11.72)$ & $(11.17,11.27,11.71)$ & $(11.17,11.27,11.71)$ & $(11.17,11.27,11.71)$ \\
&$2.0$ & $(5.68,5.72,5.91)$ & $(5.68,5.72,5.90)$ & $(5.68,5.72,5.90)$ & $(5.68,5.72,5.90)$ & $(5.68,5.72,5.90)$ \\
\hline
\hline
 & \multicolumn{5}{c}{$n=15$} \\
\multirow{6}{*}{RR$_{2,3}-\gamma^2$} &$0.5$ & $(2.12,2.12,2.13)$ & $(2.12,2.12,2.13)$ & $(2.12,2.12,2.13)$ & $(2.12,2.12,2.13)$ & $(2.12,2.12,2.13)$ \\
&$0.7$ & $(4.12,4.08,4.19)$ & $(4.12,4.08,4.18)$ & $(4.12,4.08,4.18)$ & $(4.12,4.08,4.18)$ & $(4.12,4.08,4.18)$ \\
&$0.8$ & $(19.82,19.40,20.05)$ & $(19.83,19.41,19.99)$ & $(19.86,19.45,20.00)$ & $(19.85,19.45,20.00)$ & $(19.84,19.43,19.98)$ \\
&$1.3$ & $(9.68,9.83,10.47)$ & $(9.68,9.82,10.43)$ & $(9.68,9.82,10.42)$ & $(9.68,9.82,10.42)$ & $(9.68,9.82,10.41)$ \\
&$1.5$ & $(3.32,3.36,3.54)$ & $(3.32,3.36,3.53)$ & $(3.32,3.36,3.52)$ & $(3.32,3.36,3.52)$ & $(3.32,3.36,3.52)$ \\
&$2.0$ & $(2.13,2.14,2.18)$ & $(2.13,2.14,2.17)$ & $(2.13,2.14,2.17)$ & $(2.13,2.14,2.17)$ & $(2.13,2.14,2.17)$ \\
\hline
\multirow{6}{*}{RR$_{3,4}-\gamma^2$}  &$0.5$ & $(3.02,3.02,3.03)$ & $(3.02,3.02,3.03)$ & $(3.02,3.02,3.03)$ & $(3.02,3.02,3.03)$ & $(3.02,3.02,3.03)$ \\
&$0.7$ & $(4.02,4.02,4.10)$ & $(4.02,4.02,4.10)$ & $(4.01,4.02,4.09)$ & $(4.02,4.02,4.09)$ & $(4.01,4.02,4.09)$ \\
&$0.8$ & $(13.91,13.92,14.46)$ & $(13.90,13.92,14.44)$ & $(13.90,13.91,14.42)$ & $(13.92,13.90,14.42)$ & $(13.91,13.90,14.42)$ \\
&$1.3$ & $(9.78,9.91,10.47)$ & $(9.78,9.90,10.44)$ & $(9.78,9.90,10.43)$ & $(9.78,9.90,10.42)$ & $(9.78,9.90,10.42)$ \\
&$1.5$ & $(4.10,4.14,4.29)$ & $(4.10,4.13,4.28)$ & $(4.10,4.13,4.28)$ & $(4.10,4.13,4.28)$ & $(4.10,4.13,4.28)$ \\
&$2.0$ & $(3.09,3.10,3.13)$ & $(3.09,3.10,3.13)$ & $(3.09,3.10,3.13)$ & $(3.09,3.10,3.13)$ & $(3.09,3.10,3.13)$ \\
\hline
\multirow{6}{*}{RR$_{4,5}-\gamma^2$} &$0.5$ & $(4.01,4.01,4.01)$ & $(4.01,4.01,4.01)$ & $(4.01,4.01,4.01)$ & $(4.01,4.01,4.01)$ & $(4.01,4.01,4.01)$ \\
&$0.7$ & $(4.58,4.59,4.64)$ & $(4.58,4.59,4.64)$ & $(4.58,4.59,4.64)$ & $(4.58,4.59,4.64)$ & $(4.58,4.59,4.64)$ \\
&$0.8$ & $(12.09,12.14,12.60)$ & $(12.09,12.14,12.57)$ & $(12.09,12.14,12.57)$ & $(12.09,12.14,12.56)$ & $(12.09,12.14,12.56)$ \\
&$1.3$ & $(10.37,10.49,11.02)$ & $(10.37,10.48,10.99)$ & $(10.37,10.48,10.98)$ & $(10.37,10.48,10.98)$ & $(10.37,10.48,10.98)$ \\
&$1.5$ & $(4.97,5.00,5.15)$ & $(4.97,5.00,5.14)$ & $(4.97,5.00,5.14)$ & $(4.97,5.00,5.13)$ & $(4.97,5.00,5.13)$ \\
&$2.0$ & $(4.07,4.08,4.10)$ & $(4.07,4.08,4.10)$ & $(4.07,4.08,4.10)$ & $(4.07,4.08,4.10)$ & $(4.07,4.08,4.10)$ \\
         \hline
    \end{tabular}}
    \caption{The $ARL$ values of the RR$_{r,s}-\gamma^2$ control charts in the 
   presence of measurement errors for $\gamma_0=0.05$ (left side), $\gamma_0=0.1$ (middle) and $\gamma_0=0.2$ (right side), and
   for different values of $m$, $\tau$, $n$, $\eta=0.28$, $\theta=0.05$, $B=1$.}  
  \label{tab:m}
  \end{table}

%%%Tabl8%%%%
\begin{table}
\centering
\begin{tabular}{ccccc}
\hline
 $i$ &  $\bar{X}^*_i $ & $S^*_i$  &$\hat{\gamma}$& $\hat{\gamma}^{*2}$  \\
\hline
 1&906.4&476.0 &0.525 &  0.27563\\
 2&805.1&493.9 &0.614 &  0.37700\\  
 3 &1187.2&1105.9&0.932&   0.86862  \\
 4&663.4&304.8 &0.459 &  0.21068  \\
 5&1012.1&367.4 &0.363 &  0.13177  \\
 6&863.2&350.4 &0.406 &  0.16484 \\
 7&1561.0&1562.2 &1.058 &1.11936\\
 8&697.1&253.2 &0.363 &  0.13177 \\
 9&1024.6&120.9 &0.118 &  0.01392 \\
10&355.3&235.2 &0.662 &  0.43824 \\
11&485.6&106.5 &0.219 &  0.04796   \\
12&1224.3&915.4 &0.748 &  0.55950 \\
13&1365.0&1051.6 &0.770 &  0.59290   \\
14&704.0&449.7 &0.639 &  0.40832   \\
15&1584.7&1050.8 &0.663 &  0.43957   \\
16&1130.0&680.6 &0.602 &  0.36240   \\
17&824.7&393.5 &0.477 &  0.22753   \\
18&921.2&391.6 &0.425 &  0.18062    \\
19&870.3&730.0 &0.839 &  0.70392  \\
20&1068.3&150.8 &0.141 &  0.01988    \\
\hline
\end{tabular}
 \caption{Illustrative example of Phase II dataset.}
  \label{tab:data}
\end{table}

\end{document}